\documentclass[usenatbib,usegraphicx,useAMS]{mn2e}

\usepackage{natbib}
\usepackage{bez123}
\usepackage{calc}
\usepackage{curves}
\usepackage{ebezier}
\usepackage{epic}
\usepackage{eepic}
\usepackage{latexpix}
\usepackage{multiply}
\usepackage{pgf}
\usepackage{rotating}
\usepackage{xcolor}
\usepackage{url}
\usepackage{multirow}
\usepackage{times}
\usepackage{amsmath}
\usepackage{amssymb}
\usepackage{bbold}
\usepackage{float}

\def\ltsim{\lower.5ex\hbox{$\; \buildrel < \over \sim \;$}}
\def\gtsim{\lower.5ex\hbox{$\; \buildrel > \over \sim \;$}}
\def\ltsim{\lower.5ex\hbox{$\; \buildrel < \over \sim \;$}}
\def\gtsim{\lower.5ex\hbox{$\; \buildrel > \over \sim \;$}}

\newcommand{\mbi}[1]{\mbox{\boldmath$#1$}}

\newcommand{\mat}[1]{\mbox{\rm\bf #1}}
\newcommand{\lsim}[1]{\mbox{${\,\hbox{\hbox{$ < $}\kern -0.8em \lower 1.0ex\hbox{$\sim$}}\,}$}}
\newcommand{\gsim}[1]{\mbox{${\,\hbox{\hbox{$ > $}\kern -0.8em \lower 1.0ex\hbox{$\sim$}}\,}$}}

\newcommand{\dd}{{\rm d}}

\newcommand{\cc}{{\rm c}}

\def\etal{{\it et al.\ }}

\def\beqn{\vspace{2mm}
\begin{eqnarray}} 
\def\eeqn{\vspace{2mm} 
\end{eqnarray}}

 \def\la{\langle}

\newcommand{\be}{\begin{equation}}
\newcommand{\ee}{\end{equation}}
\newcommand{\ba}{\begin{eqnarray}}
\newcommand{\ea}{\end{eqnarray}}
\newcommand{\brr}{\begin{array}}
 
\newcommand{\err}{\end{array}}
\newcommand{\bc}{\begin{center}}
\newcommand{\ec}{\end{center}}

\begin{document}
\title[Multiscale Inference from the Ly$\alpha$ Forest]{Multiscale Inference of Matter Fields and Baryon Acoustic Oscillations from the Ly$\alpha$ Forest}

\author[Kitaura \etal]{Francisco-Shu Kitaura$^{1,2}$\thanks{E-mail: francisco.shukitaura@sns.it, kitaura@usm.lmu.de}, Simona Gallerani$^{3}$ and Andrea Ferrara$^{1}$\\
 $^{1}$ SNS, Scuola Normale Superiore di Pisa, Piazza dei Cavalieri, 7 -- 56126 Pisa -- Italy \\
 $^{2}$ LMU, Ludwig-Maximilians Universit\"at M\"unchen, Scheinerstr. 1 -- D-81679 Munich -- Germany \\
 $^{3}$ INAF-Osservatorio Astronomico di Roma, via di Frascati, 33 -- 00040 Monteporzio Catone -- Italy 
}

\maketitle

\begin{abstract}
{ We present a novel Bayesian method for the  joint reconstruction of cosmological matter density fields, peculiar velocities and power-spectra in the quasi-nonlinear regime.  We study its applicability to the  Ly$\alpha$ forest based on multiple quasar absorption  spectra.  } Our approach to this problem includes a multiscale, nonlinear, two-step scheme since the statistics describing the matter distribution depends on scale, being strongly non-Gaussian on small scales ($<$ 0.1 $h^{-1}$ Mpc) and closely lognormal on scales $\gsim110\,h^{-1}$ Mpc. The first step consists on performing 1D highly resolved matter density reconstructions along the line-of-sight towards $z \sim$2--3 quasars based on an arbitrary non-Gaussian univariate model for matter statistics.  The second step consists on { Gibbs-sampling based on conditional PDFs. The matter density field is sampled in { real-space} with Hamiltonian-sampling using the Poisson/Gamma-lognormal model, while redshift distortions are corrected with linear Lagrangian perturbation theory. The power-spectrum of the lognormal transformed variable which is Gaussian distributed (and thus close to the linear regime) can consistently be sampled with the inverse Gamma distribution function.  We test our method through numerical N-body simulations with a computational volume large enough ($> 1 \,h^{-3}\,{\rm Gpc}^3$) to show that the linear power-spectra are nicely recovered over scales larger than  $\gsim120\,h^{-1}$ Mpc, i.e. the relevant range where features imprinted by the baryon-acoustics oscillations (BAOs) appear. }
\end{abstract}


\section{Introduction}
In the current cosmological picture, structures in the Universe have grown from tiny fluctuations through gravitational clustering. Nonlinear processes of structure formation destroy the information about the origin of our Universe. This information is, however, still encoded in the large scales in which structures are close to the linear regime. In particular baryon-photon plasma oscillations can be detected in the large-scale structure (LSS) as remnants of the early Universe. Their characteristic scale is measurable as an oscillatory pattern in the matter power-spectrum \citep[see e.~g.~][]{1998ApJ...496..605E} which  evolves in time in such a way that they can be used as standard rulers and to constrain cosmological parameters  \citep[see e.~g.~][]{2007AAS...210.4701E}. They can also be used to study  dark energy evolution \citep[see e.~g.~][]{2006ApJ...647....1W}.
For these reasons it is especially interesting to have measurements of baryon-acoustic oscillations (BAOs) at different redshifts and with different matter tracers confirming the same underlying physics. Therefore many efforts have been made to detect BAOs not only from the  Cosmic Microwave Background \citep[see e.~g.~][]{2003ApJS..148..135H}, but also from galaxy redshift surveys at low redshifts \citep[see][]{2005ApJ...633..560E,2006A&A...449..891H,2007MNRAS.381.1053P}, and even from photometric redshifts \citep[see][]{2007MNRAS.374.1527B}.
Moreover, using the luminous distribution of galaxies as large-scale tracers becomes extremely expensive as larger volumes need to be surveyed. 

The neutral hydrogen of the intergalactic medium (IGM) represents an appealing alternative as pointed out by several groups \citep[see e.~g.~][]{2007PhRvD..76f3009M,slosar09}. The detection of BAOs from the indirect measurements of the large-scale structure of the IGM would provide a complementary study exploiting a completely different matter tracer. Moreover, it would scan a different redshift range at which structures are closer to linear regime and thus less information loss on the cosmic initial conditions has occurred. However, many complications arise when trying to perform such a study.
Extremely luminous sources (usually quasars, but more recently also Gamma Ray Bursts) are required to shine through the IGM as distant lighthouses to allow the detection of its absorption features. The ultraviolet radiation emitted by a quasar suffers resonant Ly$\alpha$ scattering as it propagates through the intergalactic neutral hydrogen. In this process, photons are removed from the line-of-sight resulting in an attenuation of the source flux, the so-called Gunn-Peterson effect. The measurement of multiple quasar absorption sight-line spectra traces the LSS. 

The first problem consists on translating the observed flux of a quasar absorption spectrum into the underlying density field. The explicit flux-density relationship is complex.  One of the pioneering works to invert this relation was done by \citet[][]{1999MNRAS.303..179N}. Here an explicit relation between the HI optical depth, which is related to the observed flux, and the matter field is inverted using the Richardson-Lucy deconvolution algorithm \citep[][]{1972JOSA...62...55R,1974AJ.....79..745L}. In this approach a particular equation of state for the IGM has to be assumed which implies the knowledge of the gas thermal and ionization history. 
To alleviate these problems, we {propose to follow the works of local mapping methods proposed to Gaussianize cosmic fields  \citep[see][]{1992MNRAS.254..315W} and actually applied to power-spectrum estimation from the Ly$\alpha$ forest \citep[see][]{1998ApJ...495...44C,1999ApJ...520....1C}.  We note that this approach is tightly related to the biasing studies from galaxy surveys proposed by \citet[][]{2000ApJ...540...62S,2004ApJ...602...26S}. 
In particular we rely on the recently developed} 1D technique that recovers the nonlinear density field from Ly$\alpha$ data without information on the equation of state, the thermal history or the ionization level of the IGM \citep[][]{simo}. The strong correlation we found between the flux and the matter density enables one to establish a statistical one-to-one relation between the probability density of the flux and the matter one. This approach reduces all the assumptions to the knowledge of the matter statistics which is well constrained by N-body simulations. It permits us to deal with strongly skewed matter probability distribution functions (PDFs) which apply at very small scales ($<1$ $h^{-1}$ Mpc) corresponding to the Jeans scale of the IGM. This method is especially well suited for the large-scale structure analysis as it { makes a minimum number of assumptions}
and {is} computationally very efficient. {As we will show the main sources of uncertainties in the density field along the line-of-sight with our approach will come from the peculiar motions neglecting errors in the determination of the continuum flux.}

The second problem affecting power-spectrum extraction from a set of multiple quasar sight lines comes from the window function. A number of well established techniques performing a similar task from galaxy redshift surveys are available \citep[see e.~g.~][]{FELDMAN1994,TEGMARK1995,HAMILTON1997A,YAMAMOTO2003,PERCIVAL2004,PERCIVAL2005}. The main problem which arises when doing such a study comes from the aliasing introduced by the mask and selection effects of the survey. Recently, great effort has been put in designing surveys which have {\it well behaved} masks  to minimize these effects, as the 2dF Galaxy Redshift Survey\footnote{http://www.mso.anu.edu.au/2dFGRS/} \citep[][]{2003astro.ph..6581C}
and the Sloan Digital Sky Survey (SDSS)\footnote{http://www.sdss.org/} \citep[][]{2009ApJS..182..543A}.
This strategy is very limited in the case of the Lyman alpha forest. The nature of the observable, namely multiple line-of-sight quasar spectra, produces a complex 3D completeness with many unobserved regions in-between the spectra. This effect is unavoidable as quasars are sparsely distributed in space. {An alternative proposed by \citet[][]{slosar09} is to work with cross-power spectra. }

Here we propose to extend the Bayesian nonlinear Poisson{/Gamma}-lognormal models developed in \citet[][]{kitaura_log} together with the  Gibbs and Hamiltonian sampling technique as presented by \citet[][]{kitaura,jasche_gibbs} and \citet[][]{jasche_hamil} to {measure power-spectra and} detect BAOs. This approach enables us to measure the features in three dimensional power-spectrum sampling over a large ensemble of density realizations and power-spectra. 
While the concept of data augmentation with constrained realizations is clear in the Gaussian case \citep[][]{1991ApJ...380L...5H,vdw96}, it becomes very complex for nonlinear density distributions resulting only in an approximate procedure in most of the cases \citep[][]{1998ApJ...492..439B}.
The work of \citet[][]{pichon} relies on a nonlinear least squares approach reconstruction method by \citet[][]{tarantola} {which is based on Gaussian distribution functions including a nonlinear transformation.} This method has been shown to be adequate to study the topology of the IGM \citep[][]{pichon,2008MNRAS.386..211C}. 

{
The Hamiltonian sampling technique is more general and enables one to sample nonlinear distribution functions in an exact way \citep[see][]{duane,neal1993,2008MNRAS.389.1284T,jasche_hamil}. The data augmentation step is { built-in} in this method as the posterior distribution function of the matter field given the data is fully sampled. However, as we will show here the power-spectrum cannot be extracted in a straightforward way solely with the Hamiltonian sampling scheme. 

In this work we propose a simple and efficient  approach to jointly sample power-spectra and non-Gaussian density fields.
The idea consists on using the Gaussian prior for the matter field and encoding the transformation of the non-Gaussian density field into its linear-Gaussian component in the likelihood. The advantage of this approach is that we can  model the PDF of the power-spectrum with the inverse Gamma distribution in a consistent way under the Gaussian prior assumption  \citep[see][]{kitaura}. We rely on the lognormal transformation to relate the nonlinear density field and its linear counter-part as it has been done in other works to model the IGM \citep[see][]{viel,simoorg}. Also note the recent works on the lognormal transformation and its relation to the linear field by \citet{2009ApJ...698L..90N,2011ApJ...731..116N,kit_lpt}. 
As we know how to sample the density field and the power-spectrum conditioned on each other we can apply the Gibbs-sampling scheme proposed in \citet[][]{kitaura,jasche_gibbs} extended here to non-Gaussian density fields.

We show also how to solve within this framework for redshift distortions in the quasi-nonlinear regime relying on linear Lagrangian perturbation theory \citep[][]{1970A&A.....5...84Z,1990MNRAS.242..428M}.

Finally we validate our method with numerical tests based on a large N-body simulation at redshift $z=3$ of 1.34 $h^{-1}$ Gpc side length which was performed by \citet[][]{angulo} to study the detectability of BAOs at different redshifts.
}

The rest of the paper is structured as follows. In the next section (\S \ref{sec:mult}) we summarize the multiscale approach; in \S \ref{sec:1D} we present the reconstruction along the quasar sight lines {and estimate the uncertainties including biasing, thermal broadening and peculiar motions}. 
{In  \S \ref{sec:3D} we present the Bayesian method to jointly recover the large-scale structure and its power-spectrum (with the BAO signal) correcting for redshift distortions based on a combination of the Gibbs and Hamiltonian sampling techniques. The numerical validation experiments are presented in section \S \ref{sec:num}.}  Finally, we present our summary and conclusions.

\section{Multiscale Statistical Inference}
\label{sec:mult}

The matter distribution changes as cosmic evolution triggers structure formation.  
 The fluctuations of the Universe which are closely normal distributed in the early epochs start to become non-Gaussian through  gravitational clustering at small scales. For the quasi-nonlinear regime the {lognormal distribution function} is known to give a good description. 

In this approximation {a nonlinear transformation of the overdensity field leads to a quantity which is assumed to be Gaussian distributed} \citep[][]{1991MNRAS.248....1C}. This will be crucial for our power-spectrum sampling method (see \S \ref{sec:3D}).

 This distribution can be considered to be valid at scales $\gsim1$10 $h^{-1}$ Mpc for the redshift range ($z\sim2-3$) we are interested in. However, on scales smaller than $\sim$0.1 $h^{-1}$ Mpc the matter statistics shows a stronger skewness than lognormal \citep[see the works by][]{colombi,2000ApJ...530....1M}. The lognormal prior can also be used for the intermediate scales $1-10$ $h^{-1}$ Mpc, as it fits very well the positive tail of the overdensity statistics, but it is not accurate in the underdense regions \citep[see][]{kitaura_log}. 
 
 Since the Jean's scale of the intergalactic gas as obtained by flux measurements of the Ly$\alpha$ forest is  $\lsim11$ $h^{-1}$ Mpc \citep[][]{1998MNRAS.296...44G}, an accurate reconstruction technique of the matter field along the spectra requires a precise treatment of the matter statistics on small scales. 
However, the baryon acoustic oscillations are washed out on small scales and become increasingly prominent towards large scales (scales larger than 10 $h^{-1}$ Mpc). We are thus dealing with two different problems which are defined on different scales. 

An approach trying to directly solve the full three dimensional problem would become either extremely complex or require strong approximations.  The multivariate matter PDF can be modeled in the highly non-Gaussian regime by expanding the lognormal distribution with the multivariate Edgeworth expansion \citep[][]{kitaura_skewlog} in analogy to the univariate case  \citep[see][]{colombi}.
However, such an expansion turns out to be extremely expensive and requires models for the multivariate higher order correlation functions (especially the three-point and the four-point statistics to model skewness and kurtosis), introducing hereby additional parameters.

Instead we propose in this work to radically simplify the problem by splitting it into
two characteristic scales: first the scale of the resolution of the data and second the scale of the minimum required resolution to study the physical problem of interest, in this case BAOs.

    In the first step we propose to apply a fast and efficient 1D reconstruction method along the line-of-sight (los) spectra. Note, that any (sufficiently fast) reconstruction method along the quasar los could be adopted \citep[e.g.~][]{1999MNRAS.303..179N}. We consider, however, a method which we have recently developed as especially adequate for this work \citep[see][]{simo} {following the works of local mapping methods proposed to Gaussianize cosmic fields  \citep[see][]{1992MNRAS.254..315W,1998ApJ...495...44C,1999ApJ...520....1C}. Our method is flexible to use an arbitrary univariate  matter distribution model for scales $<$ 0.1 Mpc. }
The method avoids any assumption on the thermal or ionization histories of the IGM which  may cumulatively affect the power on large scales and corrupt the BAOs analysis. 
It effectively includes all the physics linking the dark matter field to the observed flux, with the exception of peculiar motions, assuming the matter statistics to be known.

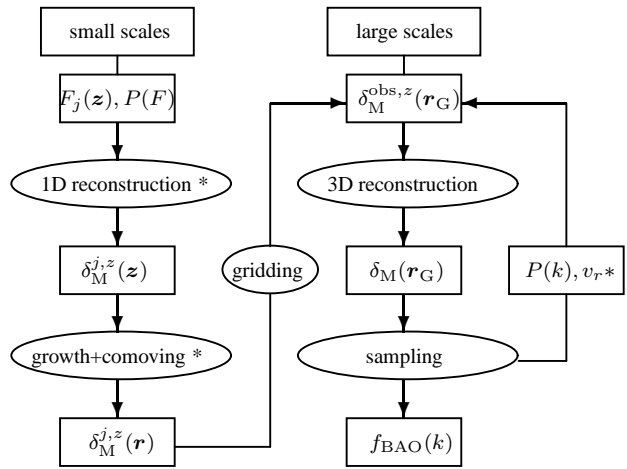
\begin{figure}
  \hspace{1.cm}
\setlength{\unitlength}{0.254mm}
\begin{picture}(349,240)(50,-250)
        \special{color rgb 0 0 0}\put(49,-61){\shortstack{$F_j(\mbi z),P(F)$}} 
        \special{color rgb 0 0 0}\allinethickness{0.254mm}\path(50,-45)(110,-45)(110,-70)(50,-70)(50,-45) 
        \special{color rgb 0 0 0}\allinethickness{0.254mm}\put(80,-70){\vector(0,-1){20}} 
        \special{color rgb 0 0 0}\put(40,-106){\shortstack{1D reconstruction *}} 
        \special{color rgb 0 0 0}\allinethickness{0.254mm}\put(82,-102){\ellipse{115}{25}} 
        \special{color rgb 0 0 0}\allinethickness{0.254mm}\put(80,-115){\vector(0,-1){20}} 
        \special{color rgb 0 0 0}\put(61,-151){\shortstack{$\delta^{j,z}_{\rm M}(\mbi z)$}} 
        \special{color rgb 0 0 0}\allinethickness{0.254mm}\path(50,-135)(110,-135)(110,-160)(50,-160)(50,-135) 
        \special{color rgb 0 0 0}\put(35,-196){\shortstack{growth+comoving *}} 
        \special{color rgb 0 0 0}\allinethickness{0.254mm}\put(82,-192){\ellipse{115}{25}} 
        \special{color rgb 0 0 0}\allinethickness{0.254mm}\put(80,-160){\vector(0,-1){20}} 
        \special{color rgb 0 0 0}\allinethickness{0.254mm}\put(80,-205){\vector(0,-1){20}} 
        \special{color rgb 0 0 0}\put(65,-241){\shortstack{$\delta^{j,z}_{\rm M}(\mbi r)$}} 
        \special{color rgb 0 0 0}\allinethickness{0.254mm}\path(50,-225)(110,-225)(110,-250)(50,-250)(50,-225) 
        \special{color rgb 0 0 0}\put(206,-61){\shortstack{$\delta^{{\rm obs},z}_{\rm M}(\mbi r_{\rm G})$}} 
        \special{color rgb 0 0 0}\allinethickness{0.254mm}\path(200,-45)(260,-45)(260,-70)(200,-70)(200,-45) 
        \special{color rgb 0 0 0}\allinethickness{0.254mm}\put(230,-70){\vector(0,-1){20}} 
        \special{color rgb 0 0 0}\put(190,-106){\shortstack{3D reconstruction}} 
        \special{color rgb 0 0 0}\allinethickness{0.254mm}\put(232,-102){\ellipse{115}{25}} 
        \special{color rgb 0 0 0}\allinethickness{0.254mm}\put(230,-115){\vector(0,-1){20}} 
        \special{color rgb 0 0 0}\put(212,-151){\shortstack{$\delta_{\rm M}(\mbi r_{\rm G})$}} 
        \special{color rgb 0 0 0}\allinethickness{0.254mm}\path(200,-135)(260,-135)(260,-160)(200,-160)(200,-135) 
        \special{color rgb 0 0 0}\put(210,-196){\shortstack{sampling}} 
        \special{color rgb 0 0 0}\allinethickness{0.254mm}\put(232,-192){\ellipse{115}{25}} 
        \special{color rgb 0 0 0}\allinethickness{0.254mm}\put(230,-160){\vector(0,-1){20}} 
        \special{color rgb 0 0 0}\allinethickness{0.254mm}\put(230,-205){\vector(0,-1){20}} 
        \special{color rgb 0 0 0}\put(212,-241){\shortstack{$f_{\rm BAO}(k)$}} 
        \special{color rgb 0 0 0}\allinethickness{0.254mm}\path(200,-225)(260,-225)(260,-250)(200,-250)(200,-225) 
        \special{color rgb 0 0 0}\allinethickness{0.254mm}\put(160,-60){\vector(1,0){40}} 
        \special{color rgb 0 0 0}\put(55,-26){\shortstack{small scales}} 
        \special{color rgb 0 0 0}\allinethickness{0.254mm}\path(40,-10)(120,-10)(120,-35)(40,-35)(40,-10) 
        \special{color rgb 0 0 0}\put(205,-26){\shortstack{large scales}} 
        \special{color rgb 0 0 0}\allinethickness{0.254mm}\path(190,-10)(270,-10)(270,-35)(190,-35)(190,-10) 
        \special{color rgb 0 0 0}\allinethickness{0.254mm}\put(160,-147){\ellipse{50}{25}} 
        \special{color rgb 0 0 0}\put(140,-151){\shortstack{gridding}} 
        \special{color rgb 0 0 0}\allinethickness{0.254mm}\path(80,-35)(80,-45) 
        \special{color rgb 0 0 0}\allinethickness{0.254mm}\path(230,-35)(230,-45) 
        \special{color rgb 0 0 0}\allinethickness{0.254mm}\path(160,-60)(160,-135) 
        \special{color rgb 0 0 0}\allinethickness{0.254mm}\path(160,-160)(160,-240) 
        \special{color rgb 0 0 0}\allinethickness{0.254mm}\path(160,-240)(110,-240) 
        \special{color rgb 0 0 0}\allinethickness{0.254mm}\path(315,-195)(290,-195) 
        \special{color rgb 0 0 0}\allinethickness{0.254mm}\path(315,-160)(315,-195) 
        \special{color rgb 0 0 0}\allinethickness{0.254mm}\path(285,-135)(345,-135)(345,-160)(285,-160)(285,-135) 
        \special{color rgb 0 0 0}\allinethickness{0.254mm}\path(315,-60)(315,-135) 
        \special{color rgb 0 0 0}\put(293.5,-151){\shortstack{$P(k),v_r*$}} 
        \special{color rgb 0 0 0}\allinethickness{0.254mm}\put(315,-60){\vector(-1,0){55}} 
        \special{color rgb 0 0 0} 
\end{picture}
\caption{Flowchart of the multiscale reconstruction method based on the \textsc{Argo}  code {\citep[first reference in][]{kitaura}}. We start with the high resolved (small scales) flux $F_j(\mbi z)$ along the los spectra $\{j\}$  from which we reconstruct the 1D matter overdensity $\delta^{j,z}_{\rm M}(\mbi z)$ {with angular and redshift coordinates $\mbi z$ including redshift distortions indicated by the superscript $z$}. Assuming linear theory we  transform the density into a comoving frame $\mbi r$ obtaining $\delta^{j,z}_{\rm M}(\mbi r)$. The set of reconstructed density lines are sorted in a lower resolved grid {(with grid positions $\mbi r_{\rm G}$)} than the binning of the spectra (large scales) leading to an incomplete matter overdensity field in redshift-space  $\delta^{{\rm obs},z}_{\rm M}(\mbi r_{\rm G})$.{ We use the Bayesian framework based on the Gibbs-sampling approach to jointly sample the matter density field $\delta_{\rm M}(\mbi r_{\rm G})$ in real-space with the Hamiltonian sampling scheme, the power-spectrum with the inverse Gamma distribution function  and correcting for redshift distortions with Lagrangian perturbation theory (see \S \ref{sec:joint})}.  {From the power-spectra} the BAO signal ($f_{\rm BAO}(k)$) can be measured. The asterisks indicate steps in which a set of cosmological parameters has to be assumed. }
\label{fig:flow}
\end{figure}

In the second step we propose to use the set of reconstructed density los spectra sorted on a lower resolution grid to recover the large-scale structure and measure the BAO signal (see flowchart at Fig.~\ref{fig:flow}). Here a Poisson/Gamma-lognormal model \citep[][]{kitaura_log} is adopted \citep[a Gaussian-lognormal model could also be adequate, see \S~\ref{sec:gauss} and ][]{pichon}. 
The gaps in-between the lines are randomly augmented by sampling the full posterior of this model with the Hamiltonian Markov Chain Monte Carlo technique, thus solving for mask induced mode-coupling in the power-spectrum \citep[][]{jasche_hamil}.  {We show how to sample the power-spectrum corresponding to the Gaussian distributed variable associated to the lognormal assumption and how to correct for linear and quasi-nonlinear redshift distortions within the Bayesian framework following the idea of \citet[][]{kitaura}. }
Details of both {1D and 3D} reconstruction methods are presented in the next sections.

\section{1D reconstruction: small scales}
\label{sec:1D}

{The purpose of this section is to make an estimation of the uncertainties in the  recovery of} the overdensity field $\delta_{\rm M}(z)$ ($\Delta_{\rm M}(z)\equiv 1+\delta_{\rm M}(z)$ with the subscript M standing either for the baryonic: B or the dark matter: D) along the los from quasar absorption spectra. 

The usual approach in signal reconstruction is to define a data model, i.~e.~the equation relating the observational data (in our case: the flux) to the desired underlying signal (in our case: the dark matter density field), and then invert this relation. In this approach, a good understanding of the data model will lead to a better estimate of the signal. This is the way \citet[][]{1999MNRAS.303..179N} propose to recover  the density field from quasar absorption spectra.

{ However, we propose to adopt a statistical model as described below \citep[see also][]{simo}.}

\begin{figure*}
\includegraphics[width=14.cm]{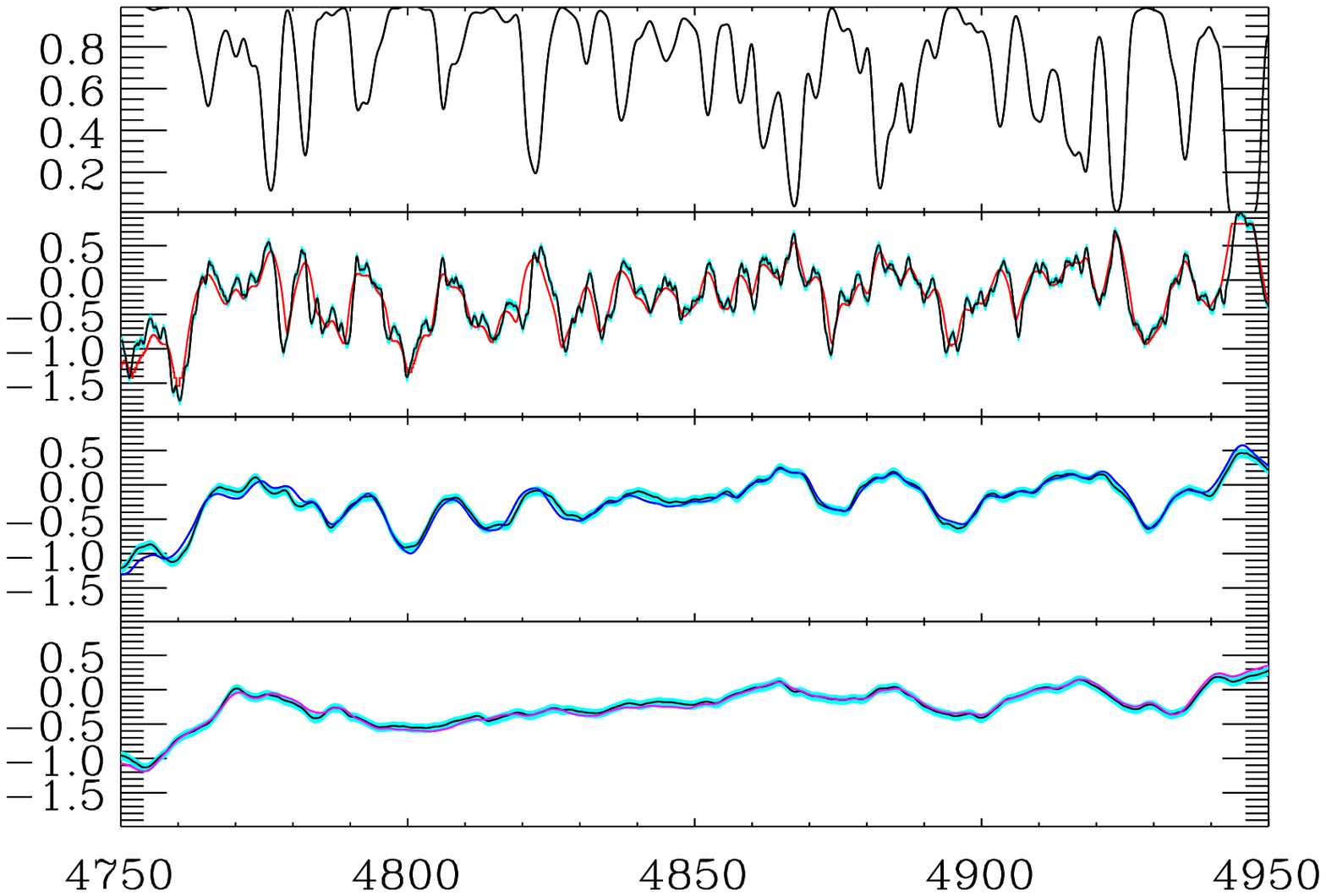}
\put(-190,0){{\large $\lambda_{\rm obs} [\AA]$}}
\put(-385,230){\rotatebox[]{90}{{\large$F$}}}
\put(-385,180){\rotatebox[]{90}{{\large$\log_{10}\Delta_{\rm B}$}}}
\put(-385,123){\rotatebox[]{90}{{\large$\log_{10}\Delta_{\rm B}$}}}
\put(-385,65){\rotatebox[]{90}{{\large$\log_{10}\Delta_{\rm B}$}}}
\put(-85,100){{\large$L_{\rm S}=5$ Mpc}}
\put(-85,40){{\large$L_{\rm S}=10$ Mpc}}
\caption{Top-most panel: example of synthetic flux quasar absorption spectrum $F$ as a function of the observed wave-length in Angstroms $\lambda_{\rm obs} [\AA]$, simulated as in \citet[][]{simo}; Top-middle panel: the original density field is shown through a black line, while the red line denotes the recovered density field through our method \citep[see][]{simo}. The cyan shaded region represents the 15\% relative error on the density field reconstruction in all the panels; Bottom-middle panel: the original density field smoothed to a scale of 5 comoving Mpc is shown through a black line, while the red line denotes the recovered density field smoothed to the same scale; Bottom-most panel; same as in the bottom-middle panel but adopting a smoothing scale of 10 comoving Mpc.}\label{fig:1drec}
\end{figure*}

\subsection{Statistical data model}

As it was already noticed by \citet[][]{1999MNRAS.303..179N} the thermal history and the ionization level of the IGM can be constrained if one knows the matter statistics. In \citet[][]{1999MNRAS.303..179N} work this is done as a consistency check once the dark matter field has already been inverted. Then the matter statistics extracted from the recovered density field is compared to the theoretical model.  

One can instead use the flux PDF, $P(F)$, and directly relate it to the matter statistics \citep[][]{simo}. 
Assuming that there is a one-to-one relation between the observational data (the flux) and the signal (the matter field) through their probability densities leads to a statistical data model:
\be
\int_{F_*}^{1}\dd F\, P(F)=\int_{0}^{\Delta^*_{\rm M}}\dd \Delta_{\rm M}\, P(\Delta_{\rm M}) \,,
\label{eq:consPDF2}
\ee
with $F_*$ and $\Delta^*_{\rm M}$ being chosen in such a way that both integrals give the same area.
This data model assigns to each flux $F_*$ a unique overdensity $\Delta^*_{\rm M}$. 
{Such a relation is however not unique due to thermal broadening, nonlocal biasing and peculiar motions. The first two effects can be shown to cancel out averaged over large scales \citep[see \S~\ref{sec:1Dnumexp} and][]{simo}. 
 Note that the peculiar velocities along the los can be estimated in an iterative fashion following \citet[][]{1999MNRAS.303..179N}. The disadvantage of this approach is that the Doppler parameter needs to be constrained in the Voigt profile to perform the deconvolution from redshift-space to real-space. We will therefore take care of the peculiar motions in the 3D reconstruction (see \S \ref{sec:3D}).}

A model for nonlocal biasing  becomes unnecessary for our large-scale structure analysis regarding the scales of interest (see discussion in \S~\ref{sec:bias}). Additional complications to biasing could arise in the presence of an inhomogeneous UV-background. Such an effect would become hard to quantify, however, we expect this contribution to be small at the redshifts of interest \citep[see][]{2005MNRAS.364.1429M}.

Due to noise in the measurements there is a minimum $F_{\rm min}$ and a maximum $F_{\rm max}$ detectable transmitted flux which are related to characteristic overdensities $\Delta^{\rm d}_{\rm M}$ and $\Delta^{\rm b}_{\rm M}$, respectively. This puts constraints on the integral limits in Eq.~(\ref{eq:consPDF2}): $F_{\rm min}<F_*<F_{\rm max}$ and hence $\Delta^{\rm b}_{\rm M}<\Delta_*<\Delta^{\rm d}_{\rm M}$.
The minimum and maximum detectable transmitted flux can be calculated from the observed noise root mean square deviation: $F_{\rm min}=\sigma_{\rm n}$;  $F_{\rm max}=1-\sigma_{\rm n}$. 
Knowing $F_{\rm min}$ and $F_{\rm max}$ we can determine $\Delta^{\rm d}_{\rm M}$ and $\Delta^{\rm b}_{\rm M}$ using the following analogous relation:
\be
\int_0^{F_{\rm min}}\dd F\, P(F)=\int_{\Delta^{\rm d}_{\rm M}}^{\infty}\dd \Delta_{\rm M} \, P(\Delta_{\rm M}) \,,
\label{eq:consPDF1}
\ee
\be
\int_{F_{\rm max}}^1\dd F\, P(F)=\int_0^{\Delta^{\rm b}_{\rm M}}\dd \Delta_{\rm M} \, P(\Delta_{\rm M}) \,.
\label{eq:consPDF3}
\ee

The problem of underestimating (overestimating) the density in regions with $\Delta_{\rm M}\geq\Delta^{\rm d}_{\rm M}$ ($\Delta_{\rm M}\leq\Delta^{\rm d}_{\rm M}$) can be easily solved in our approach by randomly augmenting the density according to the matter statistics \citep[][]{simo}. This could be important  in redshift ranges with large saturated regions estimating the power-spectrum. Note however, that in the redshift range $z=$2--3 we do not run into such problems as the saturated regions are rare ($\lsim110$\%).

 The inference of $\Delta^{*}_{\rm D}-\Delta^{\rm d}_{\rm D}-\Delta^{\rm b}_{\rm D}$ only depends on the
assumed PDF of the density field and not on any assumption concerning the IGM thermal and ionization history 
(see Eqs.~\ref{eq:consPDF2}, \ref{eq:consPDF1}, \ref{eq:consPDF3}), which are encoded in the observed PDF of the transmitted flux.

Summarizing, the method we intend to adopt to reconstruct the density
field along the los towards a quasar is the following: at
each bin with  $F_*\leq F_{\rm min}$ we assign an overdensity  $\Delta^{\rm d}_{\rm M}$, at each bin with  $F_*\geq F_{\rm max}$ we assign an overdensity  $\Delta^{\rm b}_{\rm M}$, while at each flux $F_{\rm min}<F_*<F_{\rm max}$ we associate an overdensity $\Delta^{*}_{\rm M}$ satisfying Eq.~(\ref{eq:consPDF2}).
The uncertainties in the reconstruction can be obtained by  propagating the uncertainties in the measured flux PDF and 
the assumed matter PDF with Eq.~(\ref{eq:consPDF2}).

The only parameters required in the statistical data model are those describing the matter PDF; the rest of parameters are constrained by observations. In the lognormal approximation the matter statistics is fully determined by the corresponding correlation function \citep[][]{1991MNRAS.248....1C}. The dark matter correlation function is determined by the cosmological parameters.  If one wants to recover the baryon density field one has to include a bias to transform the dark matter correlation function into the baryonic correlation function. 
As we have discussed above the lognormal approximation fails on these small scales. If one perturbs the lognormal PDF with the Edgeworth expansion, one can model the skewness and kurtosis of the matter PDF with the three point and the four point statistics which for the univariate case are simply numbers \citep[][]{colombi}. This introduces two additional parameters.
\citet[][]{2000ApJ...530....1M} present a fitting formula based on numerical N-body simulations which requires four parameters and accounts for the baryonic matter density distribution. Note, however, that implicitly more parameters flow in through the numerical calculations. In particular the initial conditions will be determined by the cosmological parameters.

\subsection{Bias relation between baryonic and dark matter}
\label{sec:bias}

{Here we describe the biasing model we are considering in the 1D numerical experiments.}
As we have discussed in the section above, we can avoid having to formulate explicit biasing relations by choosing the matter statistics corresponding to the species we want to recover. If we set the dark matter statistics in Eq.~(\ref{eq:consPDF2}) we will get an estimate for the dark matter density field.

Assuming that the low-column density Ly$\alpha$ forest is produced by
smooth fluctuations in the intergalactic medium which arise as a
result of gravitational instability we can relate the overdensity of the IGM
to the dark matter overdensity through a convolution \citep[][]{1997ApJ...479..523B}.
In Fourier space representation we have:
\be
\hat{\delta}_{\rm B} ({k}, z)  \equiv 
B(k,z) \hat{\delta}_{\rm D}({k},z)\,,
\ee
where 
$\hat{\delta}_{\rm D}({k},z)$ is the Fourier transformed dark matter
linear overdensity with the hats denoting the Fourier transformation \citep[see discussion in][]{viel}. 
Following \citet[][]{1997ApJ...479..523B} one approximates the bias by: $B(k,z) = (1 + k^2/k_J^2)^{-1}$ which depends on 
the comoving Jeans' length
\be 
k_J^{-1}(z) \equiv  H_0^{-1} \left[ {2 \gamma k_B T_0(z) \over 3 \mu m_p
\Omega_{0m} (1 + z)}\right]^{1/2} \;,
\ee
with $k_B$ being the Boltzmann constant, $T_0$ the temperature at mean
density, $\mu$ the IGM molecular weight, $\Omega_{0m}$ the
present-day matter density parameter and $\gamma$ the ratio of
specific heats.  
The squared bias gives us an estimate of the power-spectrum of the IGM $P_{\rm B}(k,z)$ related to the dark matter power-spectrum $P_{\rm D}(k,z)$:
\be
P_{\rm B}(k,z) = B^2(k,z) P_{\rm D}(k,z) \,.
\ee 
Other  more complex biasing relations have been proposed in the literature \citep[][]{1998MNRAS.296...44G,2000MNRAS.317..902N,2002MNRAS.329...37M}.

It was shown by \citet[][]{1998MNRAS.296...44G} (see Fig.~1 in that publication) that biasing affects only the very small scales: $k\gsim12$ $h$ Mpc$^{-1}$. We can thus avoid biasing in the large-scale structure analysis by choosing a box with modes smaller than $k=1$ $h$ Mpc$^{-1}$. 
Our synthetic 3D reconstructions include modes up to { $k\sim0.6$} $h$ Mpc$^{-1}$ (see \S~\ref{sec:num}). This restriction does not affect the study of BAOs. 
Therefore, from now on we will make no distinction between the baryon and dark matter overdensity field and call 
it simply the matter field: $\delta_{\rm M}\equiv\delta_{\rm B}=\delta_{\rm D}$.  
{Let us nevertheless study the impact of the physics in the density reconstruction on small scales in more detail below. }

\subsection{1D numerical experiments}
\label{sec:1Dnumexp}

\begin{figure}
\includegraphics[width=8.cm]{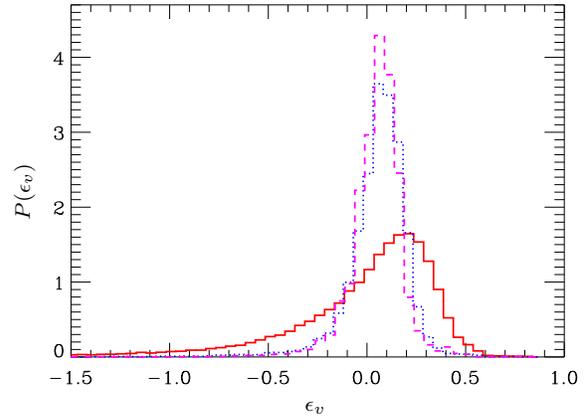}
\put(-215,80){\rotatebox[]{90}{{$P(\epsilon_v)$}}}
\put(-105,-1){{$\epsilon_v$}}
\caption{Probability distribution function of the recovered density field relative error ($\epsilon_v=(\tilde{\Delta}_{\rm B}-\Delta_{\rm B})/\Delta_{\rm B}$ with $\tilde{\Delta}_{\rm B}$ being the reconstructed density field) obtained by applying the method of density field reconstruction implemented by \citet[][]{simo} to 10 synthetic quasar absorption spectra by neglecting the peculiar velocity corrections {(i.~e. in redshift-space)}. { The solid line} represents the full resolution case, { the dotted line} shows the results for density fields smoothed to a scale of 5 comoving Mpc, { the dashed line} in the case of density fields smoothed to a scale of 10 comoving Mpc.}
\label{fig:1drec2}
\end{figure}

In this section we study the impact of saturated regions and thermal broadening together with peculiar velocities to the analysis of the cosmological large-scale structure. In order to do so, we follow the prescription by \citet[][]{simoorg} also adopted in the development of our 1D reconstruction method \citep[][]{simo}. Note that the spatial distribution of the baryonic density field and its correlation with the peculiar velocity field are taken into account adopting the formalism introduced by \citet[][]{1997ApJ...479..523B}. 
We generate with these assumptions a synthetic flux quasar absorption spectrum  (see upper-most panel in Fig.~\ref{fig:1drec}) and recover the underlying baryonic density field with our method \citep[][]{simo} (see top-middle panel in the same Fig.). We then smooth the reconstruction to $L_{\rm S}=5$ and $L_{\rm S}=10$ Mpc scales  (see bottom-middle and bottom-most panels in Fig.~\ref{fig:1drec}). In Fig.~(\ref{fig:1drec2})  the relative error of the reconstructions is represented showing that the errors are small and become symmetric distributed  at large scales.
This exercise shows us that, on small scales ($<1$ Mpc), once peculiar motions are included, the recovered density field is slightly shifted with respect to the original one. 
{At large scales}, one important point to notice is that the maximum and the minimum overdensities get below and above the saturation threshold ($-1.0\lsim1\log_{10}\Delta_{\rm B}\lsim10.5$). 

{We can therefore conclude from this study together with the ones performed in \citet[][]{simo} that saturation and thermal broadening will not affect the large-scale structure analysis. The main contribution to the uncertainties in the estimation of the density along the los with our method  come  from the peculiar motions. We will show how to correct for redshift distortions in the 3D reconstruction procedure within the Bayesian framework as proposed by \citet{kitaura}. Moreover we will demonstrate how to jointly sample the power-spectrum and the 3D matter field. } Here we are neglecting errors in the determination of the continuum flux. However, the propagation of the error in this quantity should be investigated in future works.

\begin{figure*}
\begin{tabular}{cc}
\includegraphics[width=8.cm]{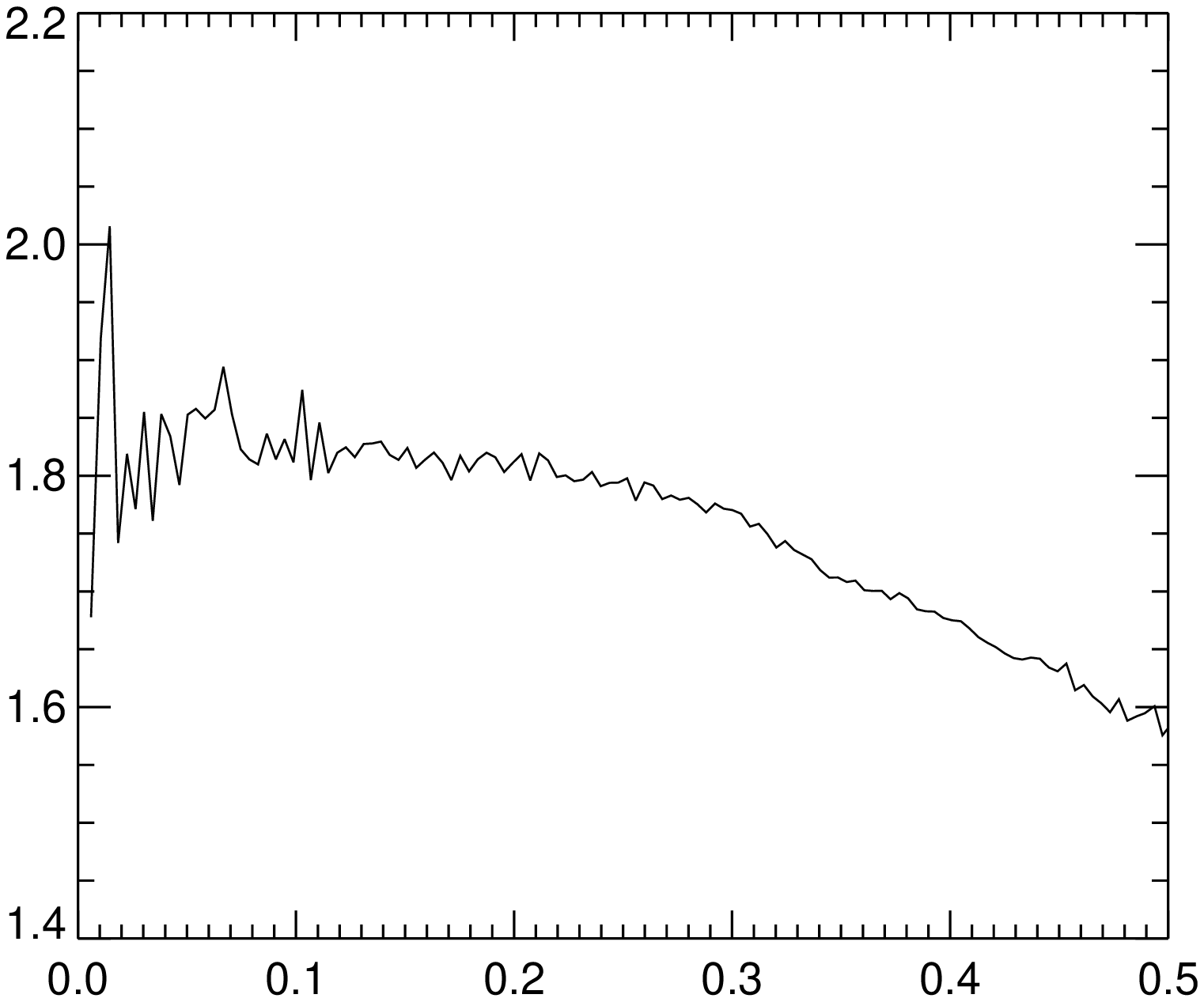}
\put(-170,160){$z=3$}
\put(-130,-5){{$k$ [$h\,$Mpc$^{-1}$]}}
\put(-240,100){\rotatebox[]{90}{{$P_z(k)/P(k)$}}}
\includegraphics[width=8.cm]{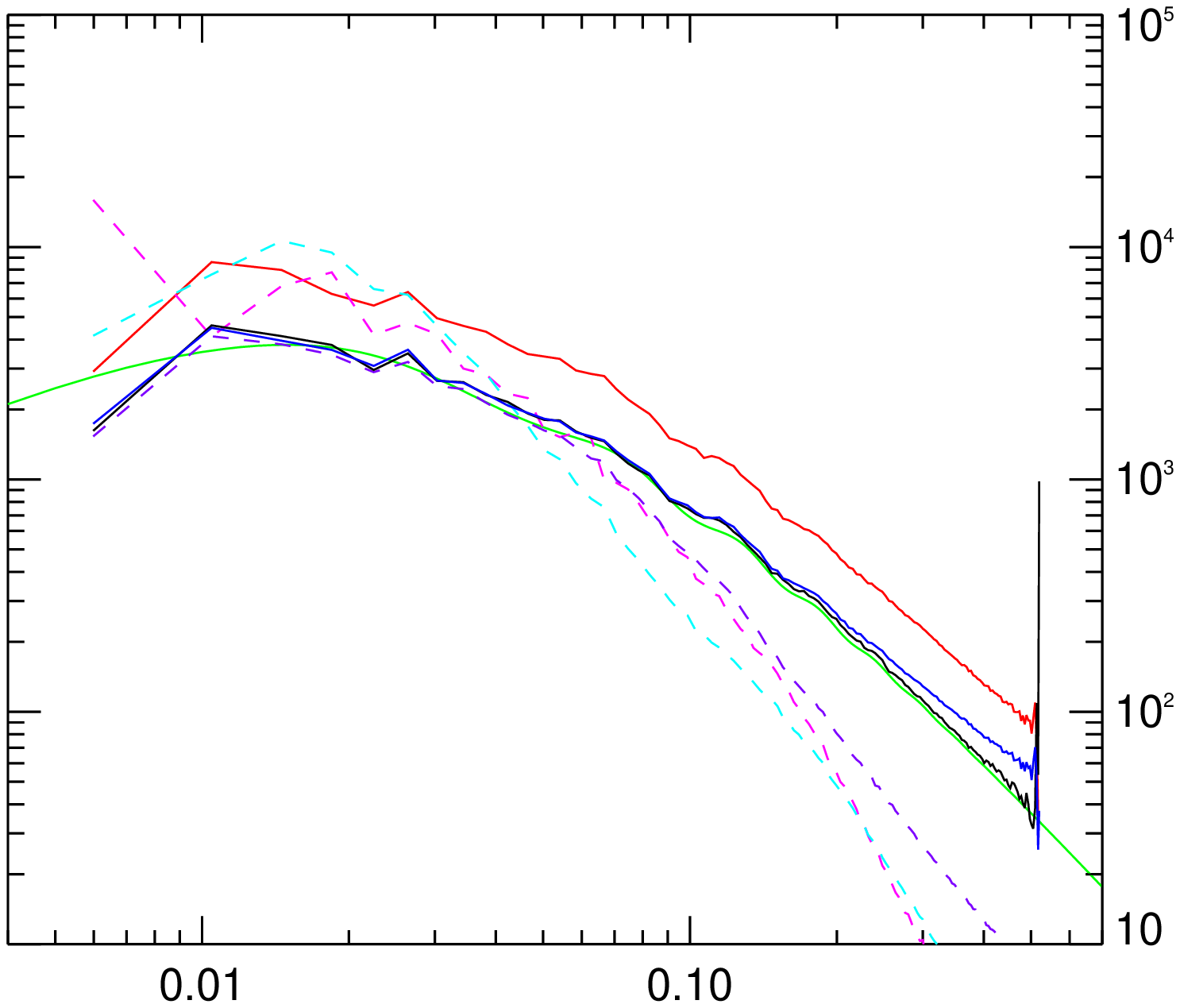}
\put(-170,160){$z=3$}
\put(0,100){\rotatebox[]{270}{{$P(k) [h^{-3}\,{\rm Mpc}^3]$}}}
\put(-130,-5){{$k$ [$h\,$Mpc$^{-1}$]}}
\put(-70,160){\color{red} z-space}
\put(-70,150){\color{magenta} it=1}
\put(-70,140){\color{cyan} it=3}
\put(-70,130){\color{violet} it=10}
\put(-70,120){\color{black} it=400}
\put(-70,110){\color{blue} r-space}
\put(-70,100){\color{green} linear theory}
\end{tabular}
\caption{Left panel: ratio between the power-spectra in redshift ($P_z(k)$) and in real-space ($P(k)$) adopting the formalism described in appendix B. Right panel: successive power-spectra showing the redshift distortions correction including the linear power-spectrum (green), the power-spectrum in redshift-space (red) and the power-spectrum in real-space (violet).}
\label{fig:ratio}
\end{figure*}

\section{3D reconstruction: large scales}
\label{sec:3D}

A set of multiple los quasar absorption spectra leads to a sparse and incomplete distribution of the dark matter density field.
In this section we present the Bayesian reconstruction method to perform an analysis of the large-scale structure based on such data following the works by \citet[][]{kitaura,kitaura_sdss,jasche_gibbs,kitaura_log,jasche_hamil}. {In this work we further extend the aforementioned techniques to perform joint reconstructions of matter fields and power-spectra including redshift distortions correction.}
In this approach a model for the signal  (here: the 3D matter field) is defined through the prior distribution function and a model for the data (here: recovered 1D density los) is defined through the likelihood.

{The density estimates along each quasar sight line lead to the gridded density field in redshift-space on a 3D mesh which we denote by $\delta^{{\rm obs,z}}_{{\rm M},i}$ with cell $i$ and a total number of $N_{\rm c}$ cells. 
 There are} two sources of uncertainty associated to this quantity: the first comes from the uncertainty in the 1D density reconstruction from the quasar spectra; the second comes from the completeness in each cell. According to our findings the errors in the matter reconstruction along the los  using our method should be {dominated by peculiar motions (neglecting errors in the determination of the continuum flux)}  after the gridding step  (see \S~\ref{sec:1D}). 
{Let us write a data model for the degraded  overdensity field in redshift-space  $\delta_{\rm M}^{{\rm obs,}z}$ 
\be
\mbi \delta_{{\rm M}}^{{\rm obs,}z}=\mat R\mbi\delta_{{\rm M}}^z+\mbi\epsilon_{z}\,,
\ee
where $\mat R$ represents the response operator and $\mbi\epsilon_{z}$ the noise. We will make the simplifying assumption that the { response} operator is a diagonal matrix given by the 3D completeness (or selection function) $R_{ij}=w_i\delta_{ij}^{\rm K}$ (where $\delta_{ij}^{\rm K}$ is the Kroenecker delta { and $w_i$ is the completeness at cell $i$).
The completeness represents the accuracy with which each cell has been sampled by the quasar sight spectra. The likelihood should therefore model the uncertainty in the density field in each cell as a function of the completeness in that cell. 
}

In the numerical experiments we will use  the Poissonian{/Gamma} likelihood as we assume that the 1D reconstruction has negligible errors and the uncertainty is only due to the incompleteness {and peculiar motions}. 
{ Note that this work could be extended within the same formalism to incorporate more complex correlations in the noise covariance using a Gaussian likelihood \citep[see appendix \ref{sec:like} and][]{pichon}.}

{We will show how to treat peculiar motions separately with a Gibbs-sampling scheme. Therefore we will first focus on the matter field reconstruction assuming that the data are transformed into real-space $\mbi\delta_{\rm M}^{{\rm obs}}$.
Let us define the statistical model for the matter field in real-space.}

\subsection{Prior for the matter statistics on large scales in real-space}

{We assume a multivariate lognormal distribution for the matter density field. The logarithm of this prior reads:
\ba
\lefteqn{-\ln(P(s|\mbi c))=}\\
&&\frac{1}{2}{\rm ln}\left(\left(2\pi\right)^{N_{\rm c}} {\rm det}(\mat S)\right) + \frac{1}{2} \mbi s^\dagger \mat S^{-1} \mbi s\nonumber\,,
\ea
with  $\mbi s=\ln(1+\mbi \delta_{\rm M})-\mbi \mu_s$, $\mbi \mu_s=\langle\ln(1+\mbi \delta_{\rm M})\rangle$ and $\mat S$ being the covariance of the field $\mbi s$. The corresponding gradient yields:
\be
-\frac{\partial\ln(P(s|\mbi c))}{\partial \mbi s}=\mat S^{-1} \mbi s \,,
\ee  
with $\mbi c$ being the set of cosmological parameters which determine the covariance $\mbi S$.}
To see how the covariance $\mat S$ is related to the Gaussian power-spectrum $P(\mbi k)$ and how to calculate the mean field $\mbi \mu_s$ see \citet[][]{1991MNRAS.248....1C} \citep[detailed derivations can be found in][]{kitaura_skewlog}.

{We should note here that the lognormal prior is a Gaussian with zero mean for the variable $\mbi s=\ln(1+\mbi \delta_{\rm M})-\mbi \mu_s$. In several works this quantity has been interpreted as an estimate of the linear component of the density field \citep[see e.g. ][]{viel,simoorg,2009ApJ...698L..90N,kit_lpt}. This motivates us to sample the power-spectrum corresponding to the variable $\mbi s$ instead of the power-spectrum of the nonlinear matter field $\mbi\delta_{\rm M}$ as we will show below.

 When generating random lognormal fields with a given power-spectrum \citep[see e.g.][]{PERCIVAL2004}  or in the case of lognormal density reconstructions \citep[see e.g.][]{kitaura_log}  one does not know the full nonlinear density field  so that the mean field $\mbi\mu_s$ cannot be directly computed from $\mbi\delta_{\rm M}$ ({ $\mbi\mu_s=\langle\ln(1+\mbi\delta_{\rm M})\rangle$}) but it is calculated from the lognormal correlation function \citep[see][]{1991MNRAS.248....1C}. However, as we want to sample over the power-spectrum we should not keep any dependence on the theoretical model of the power-spectrum. 
Actually one can demonstrate that the mean field $\mbi \mu_s$ is fully determined by the Gaussian variable $\mbi s$ by imposing fair samples of the density field ($\langle\mbi\delta_{\rm M}\rangle=0$)
\be
\mu_s=-\ln\left(\langle\exp(\mbi s)\rangle\right)\,,
\ee
(for a demonstration see appendix \ref{sec:mu}).
}

\begin{figure*}
\begin{tabular}{cc}
\includegraphics[width=6.cm]{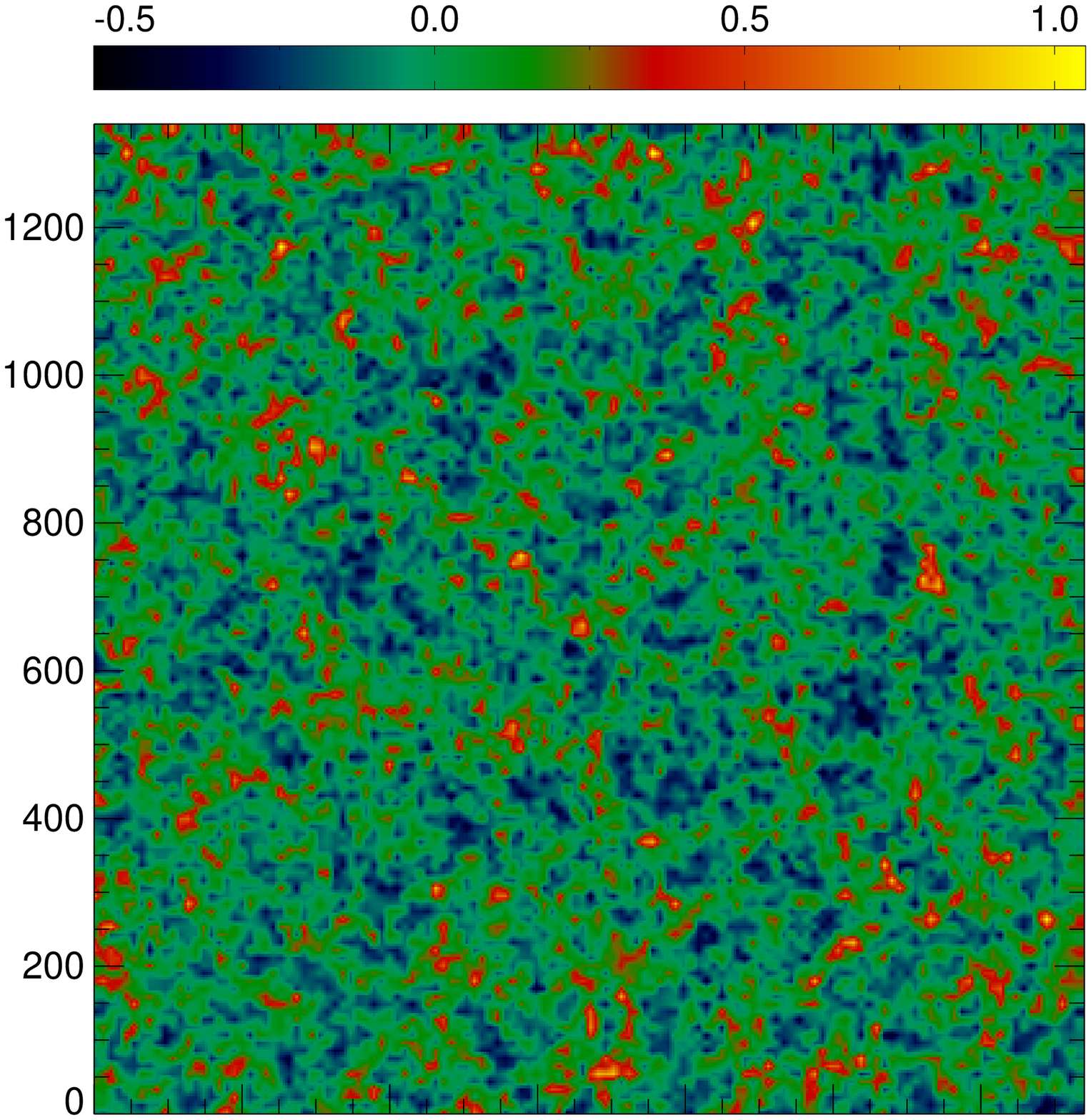}
\hspace{-.6cm}
\put(-165,117){\rotatebox[]{90}{{$Z$ [$h^{-1}\,$Mpc]}}}
\includegraphics[width=6.cm]{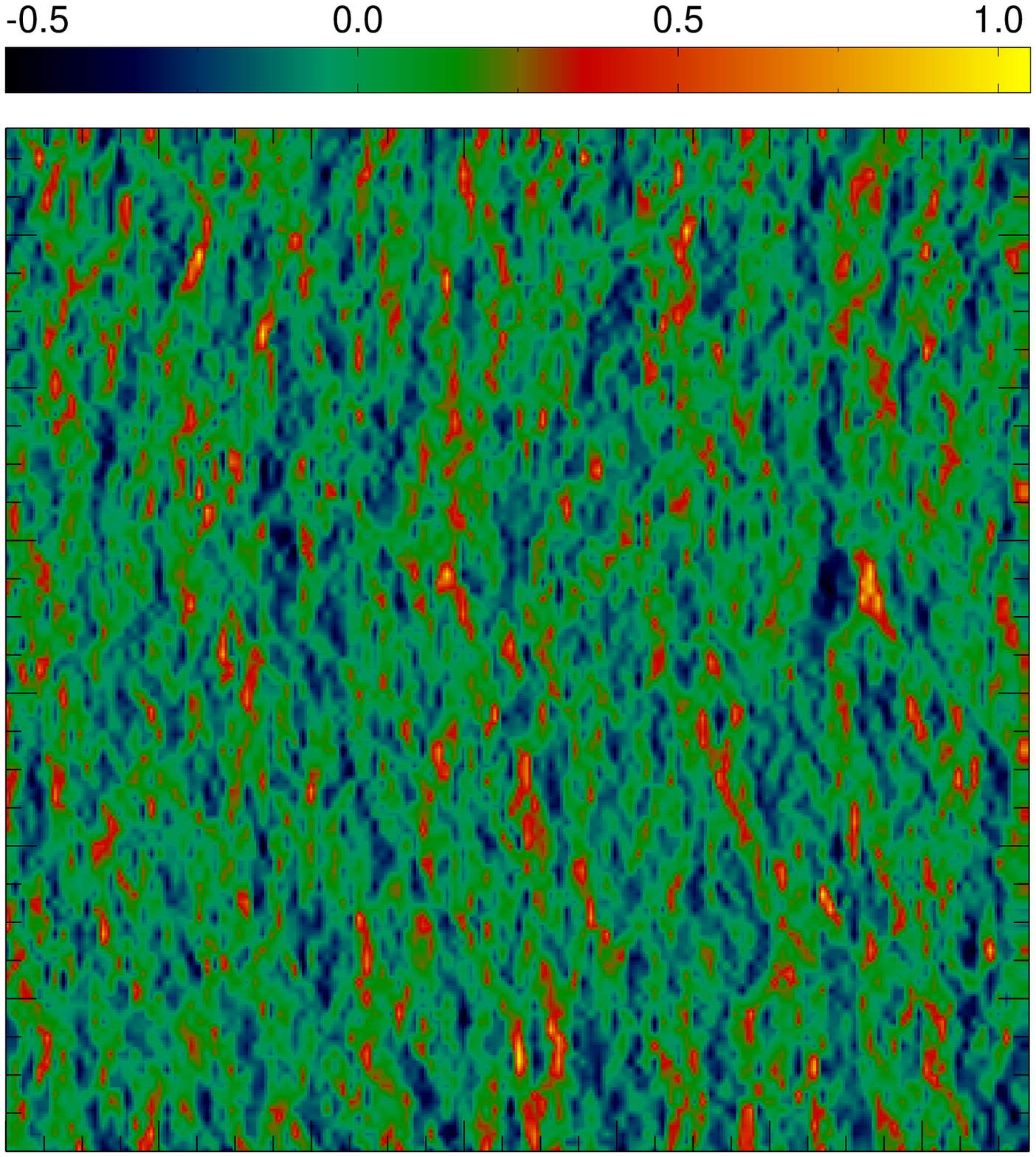}
\hspace{-.7cm}
\includegraphics[width=6.cm]{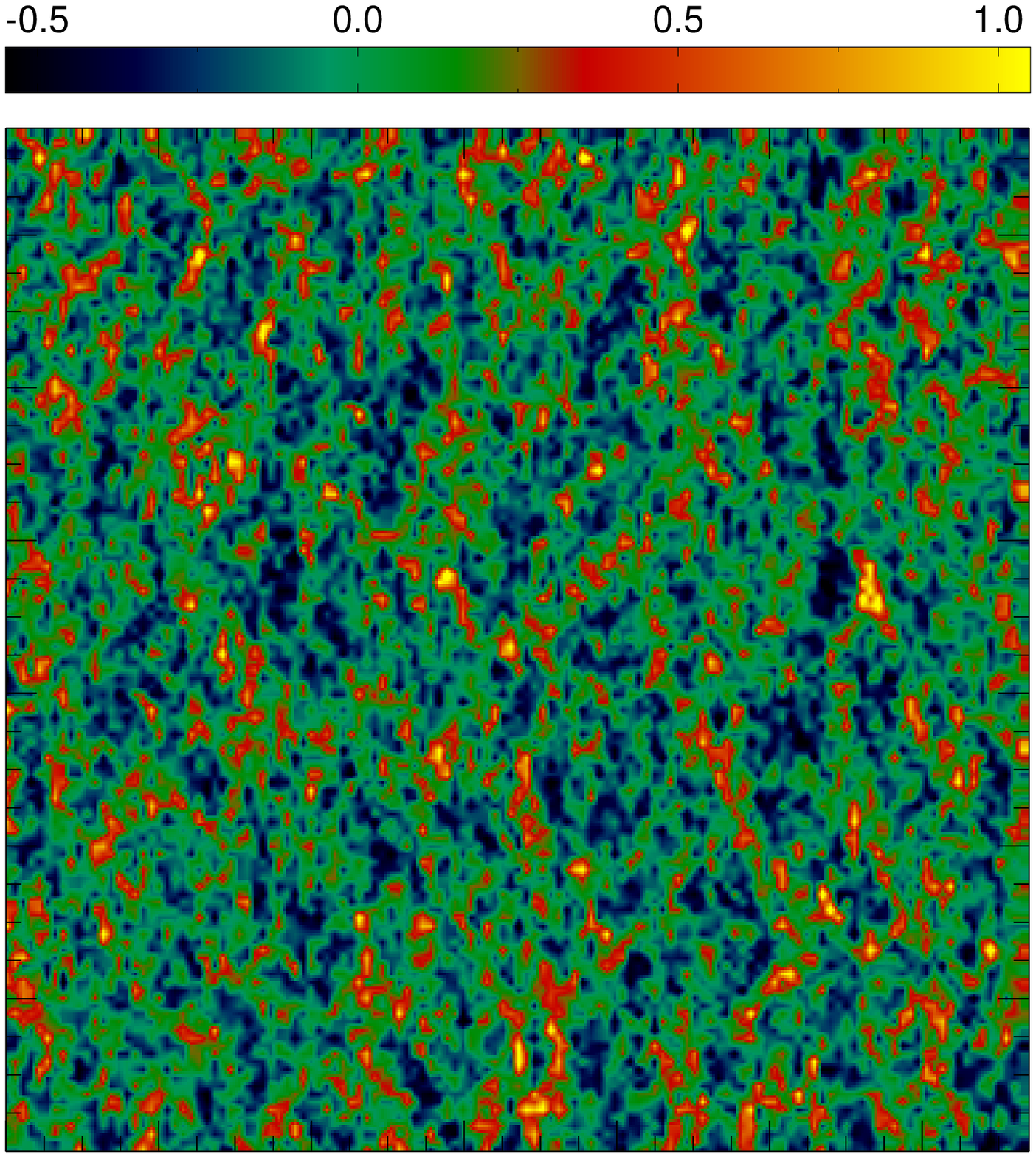}
\vspace{-2.3cm}
\\
\hspace{-0.15cm}
\includegraphics[width=6.cm]{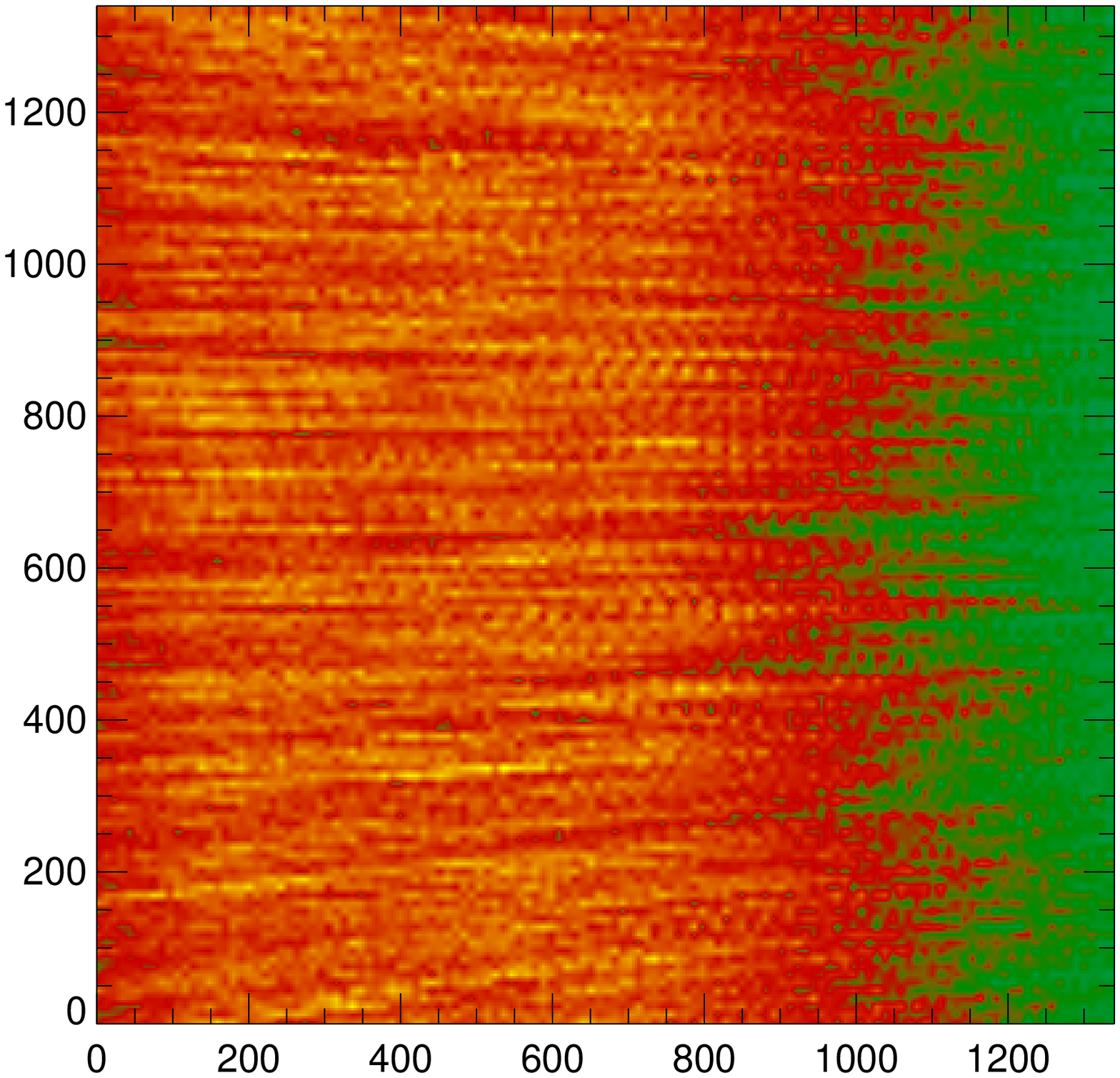}
\hspace{-.6cm}
\put(-165,115){\rotatebox[]{90}{{$Z$ [$h^{-1}\,$Mpc]}}}
\put(-90,15){{$X$ [$h^{-1}\,$Mpc]}}
\includegraphics[width=6.cm]{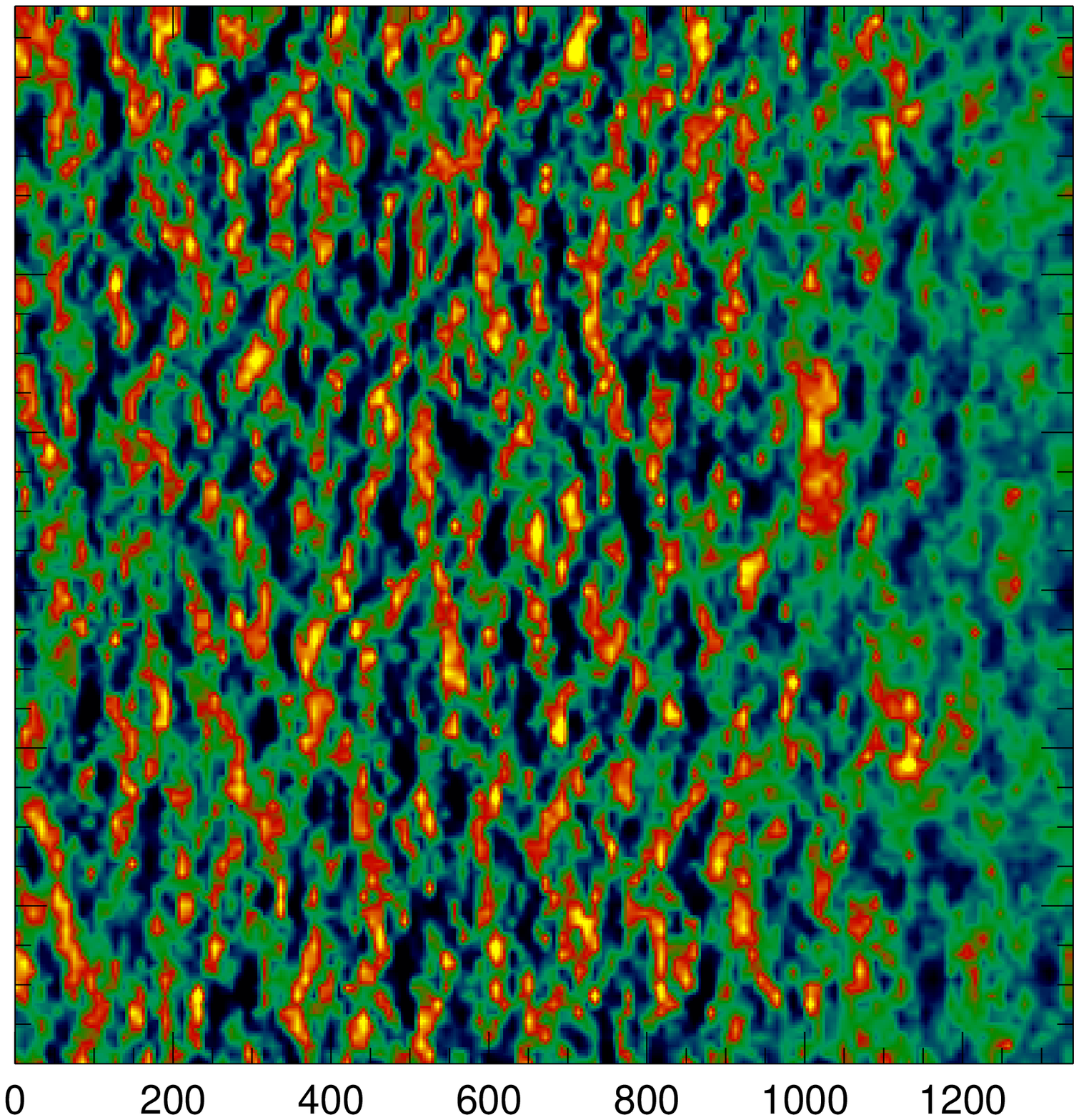}
\put(-105,15){{$X$ [$h^{-1}\,$Mpc]}}
\hspace{-.6cm}
\includegraphics[width=6.cm]{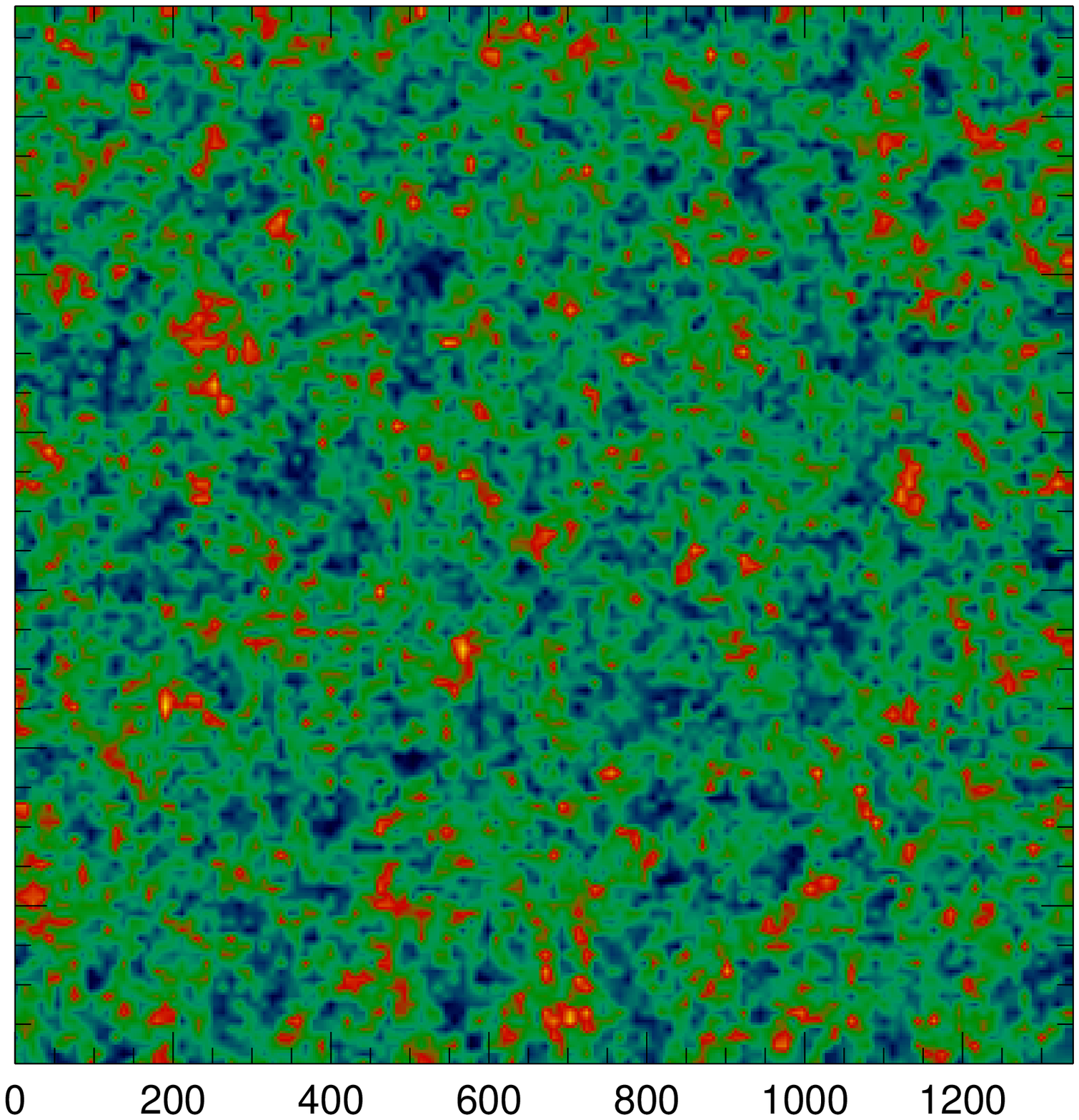}
\put(-110,15){{$X$ [$h^{-1}\,$Mpc]}}
\end{tabular}
\caption{\label{fig:deltaz} Slices of $\sim$10  $h^{-1}$Mpc width averaged over 10 neighbouring slices through the center of the box with 1.34 $h^{-1}$ Gpc side length ($Y=0$) showing upper left panel: the density field in real-space { $\delta_{\rm M}$}, upper middle panel: the velocity term $\delta_{v}$ multiplied by a factor of 2 for visualization purposes, the upper right panel: the matter field in redshift-space $\delta^z_{\rm M}$, lower left panel: the completeness according to the distribution of about 130000 mock quasar spectra multiplied by a factor of 3 for visualization purposes (ranging from 0 to $\sim0.33$), lower middle panel: 10th sample of our reconstruction, lower right panel: sample 400.}
\end{figure*}

\subsubsection{{Joint matter field reconstruction, power-spectrum sampling and redshift distortions correction}}
\label{sec:joint}

{
Here we propose a simple and straightforward approach to jointly sample power-spectra, non-Gaussian density fields and peculiar motions.
The idea consists on using the Gaussian prior for the matter field and encoding the transformation of the non-Gaussian density field into its linear-Gaussian component in the likelihood. The advantage of this approach is that it permits us to  model the PDF of the power-spectrum with the inverse Gamma distribution in a consistent way under the Gaussian prior assumption  \citep[see][]{kitaura,jasche_gibbs}.  We now rely on the lognormal transformation to relate the nonlinear density field and its linear counter-part as it has been done in other works to model the IGM \citep[see][]{viel,simoorg}. Also note, the recent works on the lognormal transformation by \citet{2009ApJ...698L..90N,2011ApJ...731..116N,kit_lpt}.  The validity of this formulation does not depend on how accurately the lognormal approximates the linear Gaussian field. As far as this work is concerned, we need to test whether the resulting sampled power-spectrum will resemble the actual linear one. We will restrict ourselves in this work to the large-scale structure at redshift $z=3$. Future work should be done to extend this study to other redshifts.

The PDF ${P(\mbi s, \mbi v, \mat S|\mbi d^z)}$ considered here is given by the joint PDF of the signal $\mbi s$, the peculiar velocity field $\mbi v$ and the power-spectrum (or its real-space counter part: the correlation function $\mbi S$). Notice that we use the following notation: the data vector in redshift-  and real-space is given by $\mbi d^z=\{\delta_{{\rm M},i}^{{\rm obs},z}\}$ and $\mbi d=\{\delta_{{\rm M},i}^{{\rm obs}}\}$ respectively\footnote{As we already discussed in \S \ref{sec:1D} one could alternatively correct for redshift distortions in the density along the los. Consequently one would  have the gridded data vector for the 3D inference analysis directly defined in real-space.}.
Knowing how to sample the conditional probability distributions of each variable
\ba
\hspace{2cm}{\mbi s^{(j+1)}}&\hookleftarrow& P({\mbi s}\mid\mbi v^{(j)},\mat S,\mbi d^z){,}\label{eq:sig} \\
\hspace{2cm}{\mat S^{(j+1)}}&\hookleftarrow& P({\mat S}\mid\mbi s^{(j+1)}) \label{eq:pow}{,}\\
\hspace{2cm}{\mbi v^{(j+1)}}&\hookleftarrow& P({\mbi v}\mid\mbi s^{(j+1)}) \label{eq:vel}{,}
\ea
permits one to sample the joint probability distribution using the Gibbs-sampling scheme \citep[see][]{gibbsamp,neal1993,toolsstatinf}. Note that the arrows indicate sampling from the conditional PDF.

Let us describe below how to sample each conditional probability distribution function
\begin{enumerate}
\item {\bf Matter field reconstruction} (Eq.~\ref{eq:sig})

This step is done with the Hamiltonian sampling method under the Gaussian prior assumption for the variable $\mbi s$ and encoding the lognormal transformation in the likelihood \citep[particular expressions and details to the Hamiltonian sampling technique can be found in appendix \ref{sec:like} and \ref{sec:MCMC}, also see][]{kitaura_log,jasche_hamil}.
 
\item {\bf Power-spectrum sampling} (Eq.~\ref{eq:pow})

We follow the works of \citet[][]{2004ApJ...609....1J,2004PhRvD..70h3511W,2004ApJS..155..227E,kitaura,jasche_gibbs} and sample the power-spectrum corresponding to $\mbi s$ { with a likelihood given by the inverse Gamma distribution function and Jeffreys prior}. For technical details in the implementation we refer to \citet[][]{jasche_gibbs}.

\item {\bf Redshift distortions correction} (Eq.~\ref{eq:vel})

Here we further develop the idea of incorporating the peculiar velocity estimation in a Gibbs-sampling scheme  \citep[see][]{kitaura}. We follow the works of \citet[][]{Kaiser-87} and in particular the expressions derived in linear Lagrangian perturbation theory (LPT) \citep[][]{1970A&A.....5...84Z} by \citet[][]{1990MNRAS.242..428M} who studied the formation of caustics in the Lyman alpha forest within this framework \citep[see][]{1990MNRAS.242..544M}.
Note, that higher order LPT may be applied following the works of \citet[][]{1999MNRAS.308..763M}. 
We should mention here that our problem (a set of quasar absorption spectra in redshift-space) requires a treatment which corrects for redshift distortions directly of the overdensity field, instead of the particle positions as in the case of galaxy redshift surveys \citep[for the latter case see the works of][]{1995A&A...298..643H,NB00,BEN02,LMCTBS08}.
We show in appendix \ref{sec:pecvel} that the observed overdensity field in redshift-space $\mbi\delta_{\rm M}^{{\rm obs,}z}$ as obtained after gridding the reconstructed overdensity fields along quasar sight lines is related to the same field in real-space $\mbi\delta_{\rm M}^{{\rm obs}}$ through the following expression
\be
\label{eq:zs}
\delta_{{\rm M},i}^{{\rm obs,}z}=\delta_{{\rm M},i}^{{\rm obs}}+\delta^{{\rm obs}}_{v,i}\,,
\ee
where one can define an observed peculiar overdensity field $\delta^{{\rm obs}}_v$
\be
\delta^{{\rm obs}}_{v,i}=w_i\delta_{v,i}+\epsilon_{v,i}\,,
\ee
which suffers from noise $\epsilon_{v,i}$ and selection function effects $w_i$.
The particular expression for the undegraded peculiar velocity term is given by
\be
\delta_v=-(Ha)^{-1}\nabla\cdot(\mbi v\cdot\hat{\mbi r})\hat{\mbi r}\,,
\ee
where   $H$ is the Hubble constant, $a$ is the scale factor and $\hat{\mbi r}$ is the los direction vector. 
We consider  linear theory to estimate the peculiar velocity field $\delta_{\rm M}\propto-\nabla\cdot\mbi v$.

In summary, the data vector is transformed from redshift- to real-space for each Gibbs-sampling iteration in step (iii) computing 
\be
\delta_{{\rm M},i}^{{\rm obs}}=\delta_{{\rm M},i}^{{\rm obs,}z}-\delta^{{\rm obs}}_{v,i}\,.
\ee
This relation is used for the next iteration in step (i) to compute the matter field reconstruction and subsequently obtain the power-spectrum sample in real-space (step ii).
Note that in the first iteration one must assume that the real- and redshift-spaces are equal $\delta_{{\rm M},i}^{{\rm obs}}=\delta_{{\rm M},i}^{{\rm obs,}z}$.

\end{enumerate}

}

\begin{figure*}
\begin{tabular}{cc}
\includegraphics[width=6.cm]{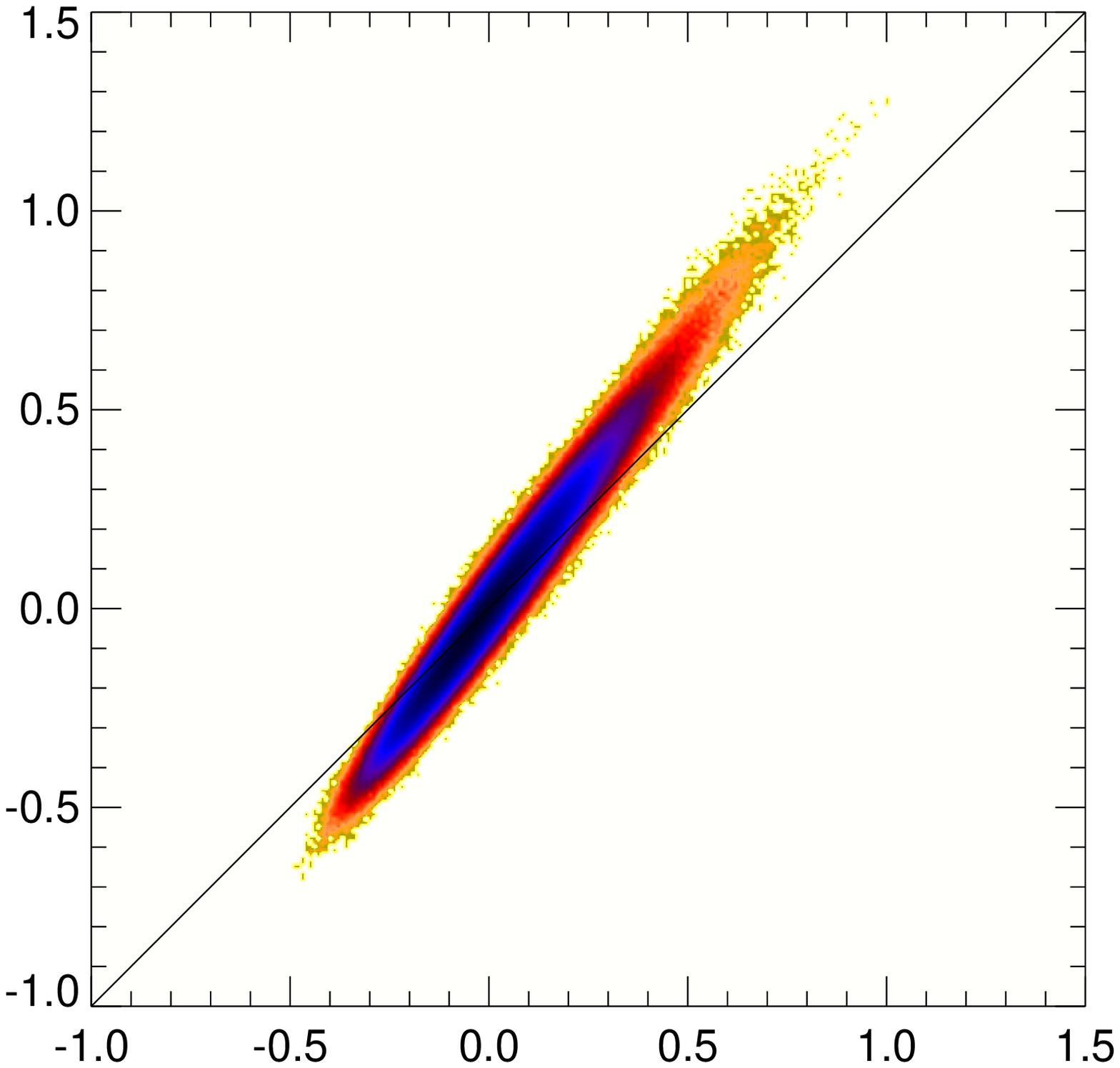}
\put(-145,140){{$\tilde{\delta}_{\rm M}=\delta^{z}_{\rm M}$}}
\put(-145,130){$z=3$}
\put(-90,40){{$r_{\rm S}=10\, h^{-1}$ Mpc}}
\put(-90,-5){{${\delta}_{\rm M}$}}
\hspace{-.5cm}
\put(-170,85){\rotatebox[]{90}{{$\tilde{\delta}_{\rm M}$}}}
\includegraphics[width=6.cm]{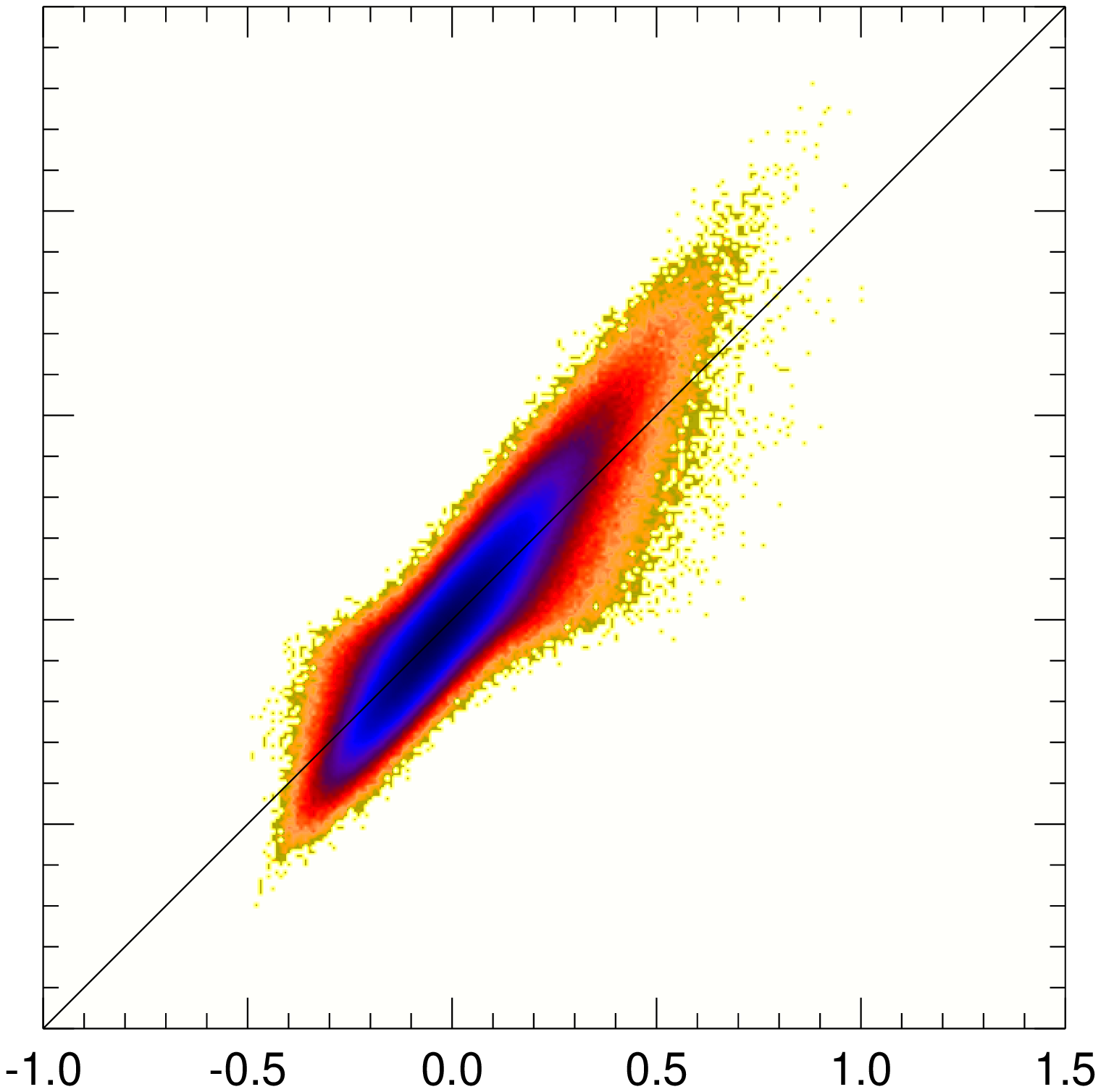}
\put(-145,140){{$\tilde{\delta}_{\rm M}=\delta^{{\rm obs},z}_{\rm M}/w$ (for $w>0$)}}
\put(-145,130){$z=3$}
\put(-90,40){{$r_{\rm S}=10\,  h^{-1}$ Mpc}}
\put(-90,-5){{${\delta}_{\rm M}$}}
\hspace{-.6cm}
\includegraphics[width=6.cm]{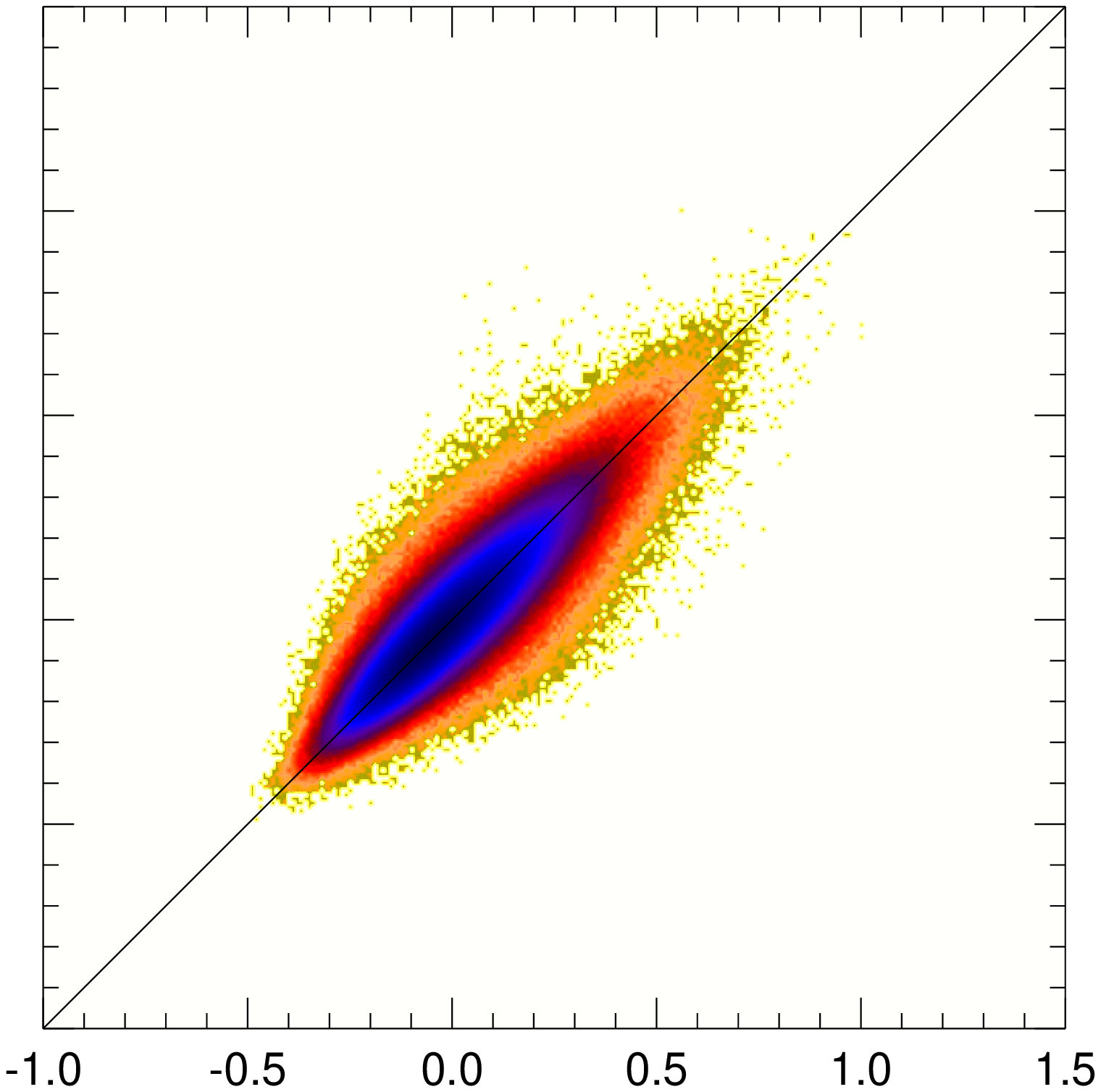}
\put(-145,140){{$\tilde{\delta}_{\rm M}=\delta^{{\rm rec}}_{\rm M}$}}
\put(-145,130){$z=3$}
\put(-90,40){{$r_{\rm S}=10\,  h^{-1}$ Mpc}}
\put(-90,-5){{${\delta}_{\rm M}$}}
\end{tabular}
\caption{\label{fig:corr} Cell-to-cell correlation between the real density field $\delta_{\rm M}$ and the one in redshift-space $\delta^{z}_{\rm M}$ (left panel), in redshift-space including mask and noise $\delta^{{\rm obs},z}_{\rm M}/w$ (the division by the mask $w$ is meant to give an estimate of the density field) (middle panel), and an arbitrary reconstructed sample $\delta^{{\rm rec}}_{\rm M}$ taking as input data $\delta^{{\rm obs},z}_{\rm M}$ (note that different converged samples give extremely similar results). The different colours indicate the number density of cells for each overdensity bin. Low density values are represented by light and   high values by dark colours.}
\end{figure*}

\subsection{3D numerical experiments}
\label{sec:num}

In this section we present  numerical tests {with the large N-body simulation provided by \citet{angulo} of 1.34 $h^{-1}$ Gpc side length} at redshift $z=3$  showing how to  solve for incompleteness in a set of multiple los density tracers { and for redshift distortions} to recover the large-scale structure, {the power-spectrum and ultimately} the BAO signal.

\subsubsection{Redshift distortions}

{
Here we validate the formalism presented in section \S \ref{sec:joint} and appendix \ref{sec:pecvel}.
In particular we check Eq.~\ref{eq:zs} by comparing the power-spectra of the resulting overdensity field in redshift-space to the actual one from the N-body simulation. We find that convolving the density field with a Gaussian kernel of smoothing length $r_{\rm S}=1.25 h^{-1}$ Mpc to estimate the velocity term gives an excellent fit to the actual ratio from the N-body simulation being close to the constant Kaiser factor ($\sim$1.8) at large scales \citep[see left panel in Fig.~\ref{fig:ratio} in this work and Fig.~7 in][]{angulo}. 

In the upper panels of Fig.~\ref{fig:deltaz} we can see slices of the N-body simulation in real-space $\delta_{\rm M}$ (left panel) with the corresponding velocity distorted component $\delta_v$ (middle panel) and the resulting overdensity field in redshift-space $\delta^z_{\rm M}$ (right panel). As the observer is located to the right of the box (at $z=0$) one can see the elongated structures along the $Z$-axis.
 
}

\subsubsection{3D mask and mock density tracers}

 The completeness of a quasar spectra distribution can be directly extracted from the data.  Only cells in which spectral lines have been detected have nonzero completeness to our purposes, otherwise the cell has not been traced by the Ly$\alpha$ forest. Either because there are no quasars in the background or that region of the sky was not observed, with the reason being irrelevant.   The completeness in observed cells can be calculated from the number density of spectra crossing that cell.
{Therefore to estimate the 3D mask (completeness) we need to simulate a distribution of quasar spectra.}

To know an approximate number of quasars in a 1.34 $h^{-1}$ Gpc comoving volume at redshift $z=$2--3 one needs to first calculate the number of  dark matter halos with masses greater than $\sim10^{12}$ solar masses.  Looking at Fig.~1 in \citet[][]{2002MNRAS.336..112M} one finds about $10^7$ objects. In order to know which fraction of halos in that mass range will host a quasar one has to consider the duty cycle.
Assuming a duty cycle of $10^{-3}-10^{-2}$ will leave us with about $10^4$--$10^5$ quasars. 
{Following previous works on forecasts for the BOSS survey we assume that we have about 130000 quasar spectra in the volume of 1.34 $h^{-1}$ Gpc side length \citep[see e.g.][]{slosar09}.}
 The mock quasars are uniformly distributed with right
ascension ($\alpha$) and declination ($\delta$) in the following
ranges: $105^{\rm o}<\alpha<270^{\rm o}$ and $-5^{\rm
  o}<\delta<70^{\rm o}$ and redshift $1.8<z<3.5$ emulating the Sloan survey. 
Then we ascribe to each mock quasar a Ly$\alpha$ forest with a maximum length of $z=0.5$ and {assume a resolution in the spectra of about 20-40 $h^{-1}$ kpc  (roughly 100-1000 bins with intervals of  $\sim$ 10-20 $h^{-1}$ Mpc comoving length)} towards the observer.  
{ The high resolution in the spectra permits us to neglect aliasing effects. }
All the redshift bins which are closer than $z=1.8$ to the observer are discarded.  We then transform each bin of the set of mock Ly$\alpha$ spectra to comoving coordinates and grid them in a box of 1.34 $h^{-1}$ Gpc and $128^3$ cells. 
The relative density of spectra determines the completeness in each cell.

\begin{figure*}
\begin{tabular}{cc}
\includegraphics[width=8.cm]{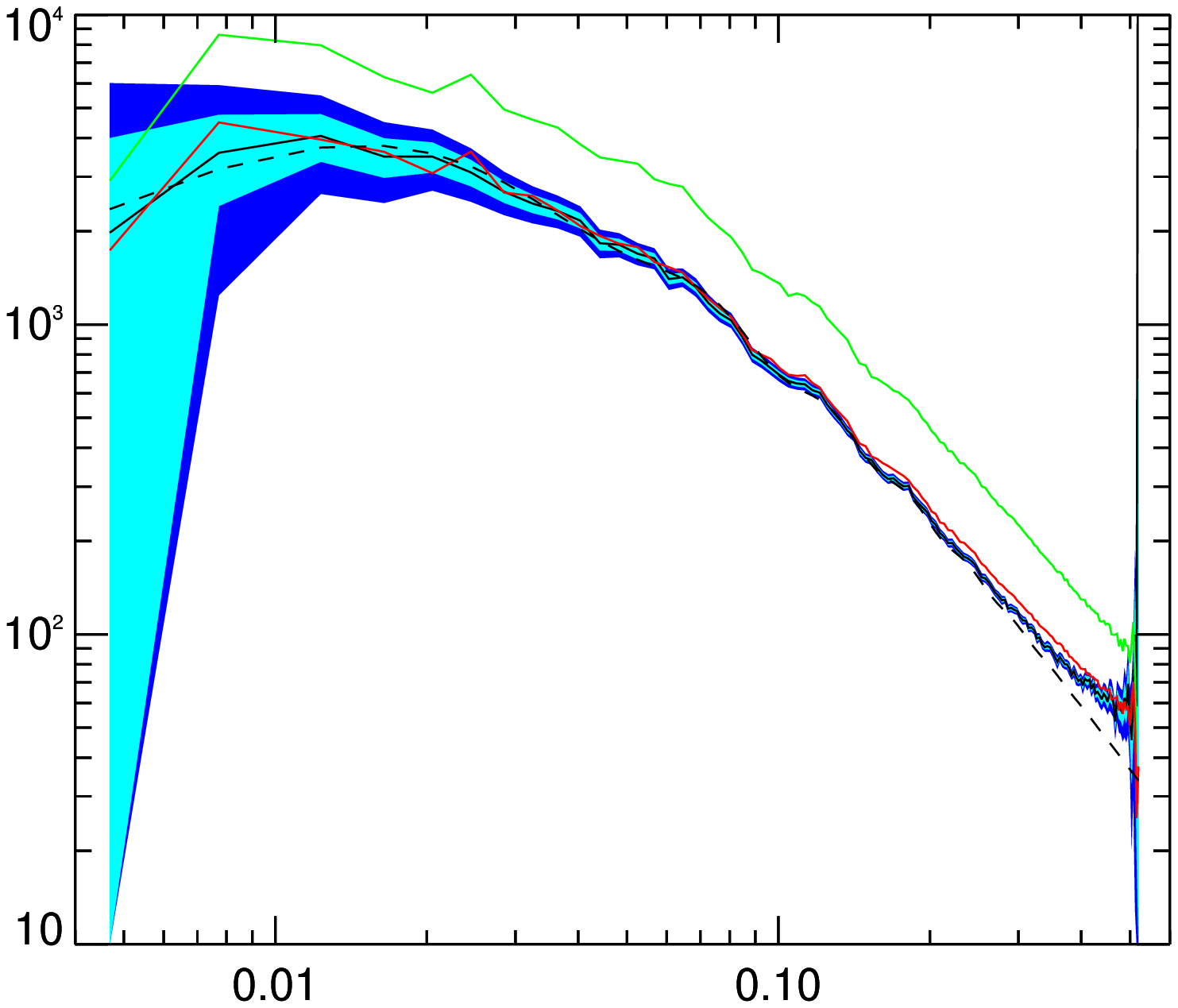}
\put(-60,150){$z=3$}
\put(-130,-5){{$k$ [$h\,$Mpc$^{-1}$]}}
\put(-240,100){\rotatebox[]{90}{{$P(k) [h^{-3}\,{\rm Mpc}^3]$}}}
\includegraphics[width=8.cm]{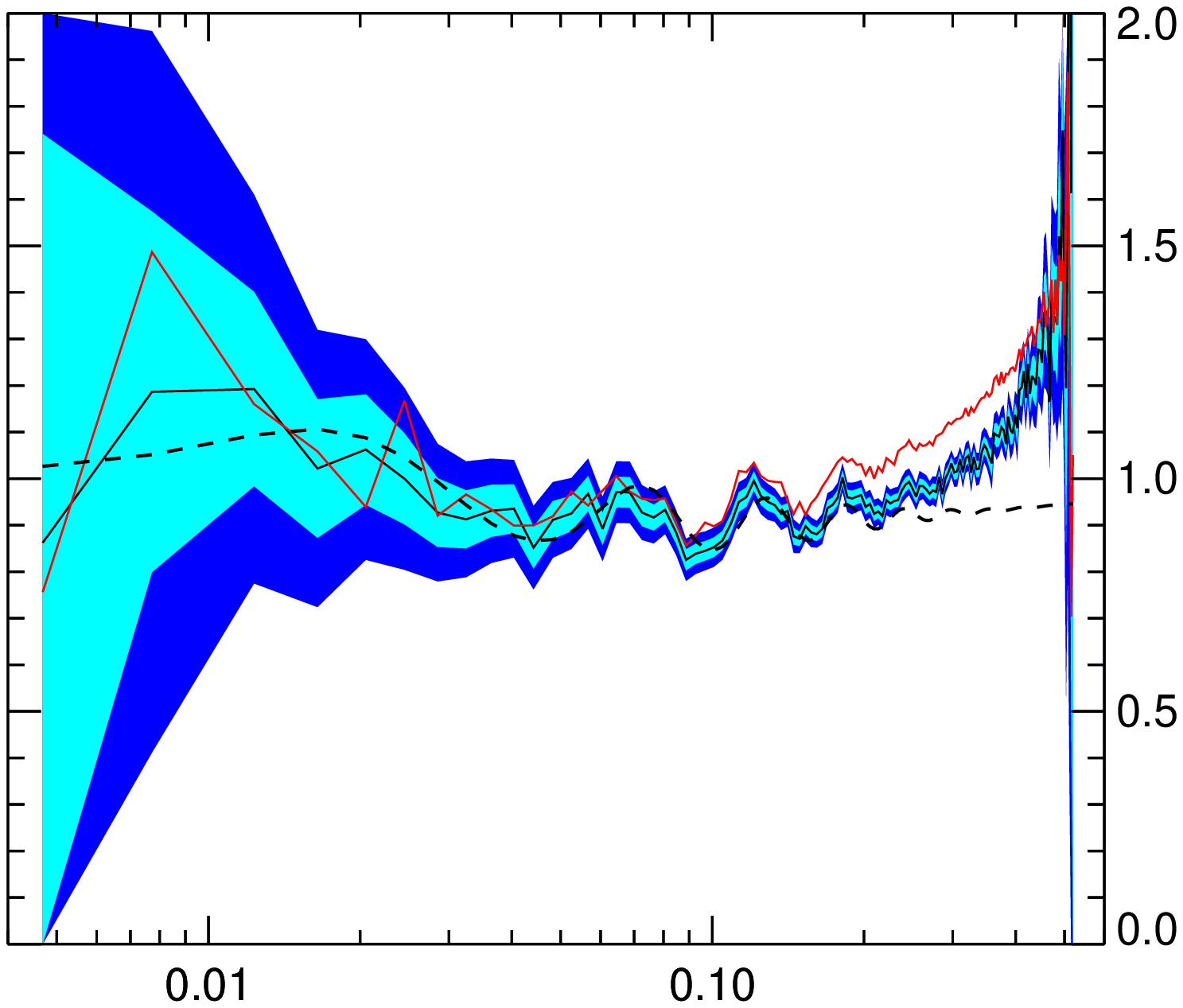}
\put(-60,150){$z=3$}
\put(0,100){\rotatebox[]{270}{{$f_{\rm BAO}(k)$}}}
\put(-130,-5){{$k$ [$h\,$Mpc$^{-1}$]}}
\end{tabular}
\caption{\label{fig:ps} Left panel: power-spectrum with 1 and 2 sigma contours (light and dark blue respectively). Black curve: mean of about 10000 samples. Red: power-spectrum of the N-body simulation in real-space. Green: same as red but in redshift-space. Dashed curve: linear theory. Right panel: same as left panel but divided by the fiducial power-spectrum without wiggles. }
\end{figure*}

{The density field is generated by gridding the dark matter particles on a $256^3$ mesh, then deconvolving with the mass-assignment kernel and finally applying a low-pass filter to map the field in Fourier-space to a lower resolution of $128^3$ \citep[see][and references there-in]{2009arXiv0901.3043J,kitaura_sdss}. Note that this procedure enables us to get nearly alias-free 3D fields. }
{ We should note, that techniques to solve for aliasing effects  on the power-spectrum \citep[see e.~g.][]{2005ApJ...620..559J} cannot be considered here as we perform a 3D analysis which includes the phase information.}

Finally, we generate a Poissonian/Gamma (see appendix \ref{sec:like}) subsample with the previously calculated completeness {from the density field in redshift-space}. 
This is equivalent to assume that the density field {in redshift-space} corresponding to each mock quasar is nearly perfectly recovered on a resolution of 10--20 $h^{-1}$ Mpc and the noise in each cell is dominated by the completeness.

\subsubsection{Reconstructions}
 
{We start running only the Hamiltonian sampling scheme together with the velocity sampling, as we find that additional power-spectrum sampling leads to extreme unphysical peculiar velocities in the burn-in period which slow down the Markov Chain or even make it crash rapidly increasing the power on large scales.
Once the the chain is nearly converged  we start sampling the power-spectrum as well. This happens only after a few hundred samples as we can see from the right panel in Fig.~\ref{fig:ps}.
We find that the errors in the determination of the peculiar velocity field, i.~e. $\epsilon_{v,i}=0$ can be neglected. Note however, that we have here full control over the model as we have tuned it with the N-body simulation. In real-data applications we should include errors which will of course increase the variance in the density fields and power-spectra. 

We also found that rarely ($<1$\% of the samples) some density samples still lead to unphysical peculiar velocities after the burn-in period. We note here that we are using Jeffreys {\it non-informative} prior for the power-spectrum and more constrained priors may be used \citep[see][]{kitaura,jasche_gibbs} which could avoid this problem. We have introduced here a rejection step for the samples which have peculiar velocity terms with $\delta_v>0.7$ as the typical values are below 0.6.

The Gibbs-sampling procedure is run for about 10000 iterations leading to an ensemble of density fields, power-spectra, and peculiar velocities.  
In the lower panels of Fig.~\ref{fig:deltaz} we can see slices of the mask (left), the 10th density sample (middle) and the 400th density sample (right).
We can appreciate in the middle plot how the regions with higher completeness (red-yellow regions in the colour code in the left panel) are especially elongated { along} the $Z$-axis showing that there is more information about the field in redshift-space which has to be transformed into real-space. The region on the right of the box is smoother as there is less information there. The sample on the right of the figure shows well balanced structures and no noticeable transition from regions with higher completeness to lower ones demonstrating that the Hamiltonian sampling technique correctly augments the field according to the model and the data. 

To see how the whole Gibbs-sampling scheme is working in a more quantitative way we show in Fig.~\ref{fig:corr} cell-to-cell correlations between the overdensity field in real-space from the N-body simulation against the field in redshift-space (left panel), the input data (middle panel) and one reconstructed sample (right panel) after convolution with a Gaussian kernel of smoothing length $r_{\rm S}=10 \, h^{-1}$ Mpc. The left panel shows in another way the same effect as we see in Fig.~\ref{fig:ratio}, namely the excess of power on large scales introduced by the linear redshift distortions which dominate at these relatively high redshifts ($z\sim3$). It is notable how the range for the overdensity field in redshift-space $\delta_{\rm M}^z$ is larger than the one for the field in real-space $\delta_{\rm M}$.  The middle plot shows the additional effect of unsampled regions and noise by having  an additional elongation around the mean density ($\delta_{\rm M}^{{\rm obs},z}\sim0$) and a larger dispersion. Note that we use the field divided by the completeness which is the flat prior estimate for the density field under selection effects \citep[see][]{kitaura_sdss}. The panel on the right demonstrates that the density samples  yield unbiased estimates of the field on scales of about $10\, h^{-1}$ Mpc. 
 Notice that the field $\delta^{\rm rec}_{\rm M}$ is the nonlinear estimate of the overdensity field computed from our samples by using the lognormal transformation $\delta^{\rm rec}_{\rm M}=\exp(s+\mu_s)-1$.

We finally show in Fig.~\ref{fig:ps} the ensemble of nearly 10000 power-spectrum samples summarized by the mean power-spectrum (black curve) together with the 1 sigma and 2 sigma countours. For comparison we also show the power-spectrum in redshift-space (green curve) and in real-space (red curve) of the N-body simulation. 
Note, that the reconstructed samples are considerably  closer to the N-body simulation in real-space than in redshift-space in terms of the power-spectra. However, we can also notice from this figure that the recovered power-spectra are closer to linear theory (dashed curve) than the red curve as we expect it from the lognormal linearization discussed in \S \ref{sec:joint}.  
} 
The mean of all samples is normalized by the assumed fiducial power-spectrum to visualize the BAO wiggles {on the right panel}.
{Note that cosmic variance is considered in a Bayesian way by sampling the inverse Gamma distribution \citep[see][]{kitaura,jasche_gibbs}.
The excess of power with respect to linear theory at small scales ($k\gsim1 0.2 \,h$Mpc$^{-1}$) can be due to various reasons: the lognormal approximation does not give perfect linearized fields, the aliasing of the gridding scheme leads to complex correlations between neighbouring cells which are not handled, or the peculiar velocity fields are not properly corrected on these scales. 
{ Also note, that we are neglecting strong correlations of the noise along quasar sight lines inside the cells which may increase the error bars, especially on small scales.}
We leave such an investigation for later work. Note that the deviation from linear theory starts at scales in which the BAO signal is washed out.
Our tests demonstrate the validity of the method to sample density fields, power-spectra and peculiar motions on large scales ($>$ 20 $h^{-1}$ Mpc) from sparse noisy data in the quasi-nonlinear regime and its use for BAO measurements. }

\section{Summary and conclusions}

{We have presented in this work a novel method for the joint reconstruction of cosmological matter fields, power-spectra and peculiar velocity fields in the quasi-nonlinear regime.  

We have applied this method to large-scale structure analysis from the Ly$\alpha$ forest based on multiple quasar absorption  spectra.}
{For this kind of studies we have proposed} to split the reconstruction problem into two steps.
The complex physical relation between the flux of quasar absorption spectra and the 1D dark matter field along the lines-of-sight is solved in the first step.
This permits us to apply simple statistical models in the second step to analyze the 3D large-scale structure. 

 Our method assumes the adequate matter statistics in each scale, being completely flexible in the choice of a non-Gaussian matter PDF in the first step  and using the multivariate lognormal model for the large-scales in the second step.

In the first step, we have based the 1D reconstruction of the matter density fields through the Ly$\alpha$ forest on a method recently introduced by \citet[][]{simo} which has the advantage of being free of any assumption on the thermal history and the ionization level of the IGM.
{In this work we have shown that saturation and thermal broadening  will not affect our large-scale structure reconstruction assuming that the errors in the determination of the continuum flux are under control. The main contribution to the uncertainties in the estimation of the density along the line-of-sight quasar spectra depends on the peculiar motions and the complex masking of a quasar absorption spectra distribution.}

In the second step we propose to apply the Bayesian {framework with the Gibbs-sampling approach to jointly sample non-Gaussian density fields, power-spectra and peculiar motions.
The idea consists on using the Gaussian prior for the matter field and encoding the transformation of the non-Gaussian density field into its linear-Gaussian component in the likelihood. The advantage of this approach is that it permits us to  model the PDF of the power-spectrum with the inverse Gamma distribution in a consistent way under the Gaussian prior assumption.  We rely on the lognormal transformation to relate the nonlinear density field and its linear counter-part.
Redshift distortions are corrected by sampling the peculiar velocity field with linear Lagrangian perturbation theory.

Finally we have validated our method with a large N-body simulation. We found that from strongly biased input data due to redshift distortions and masking our method recovers unbiased  density samples on scales of about $10\,h^{-1}$ Mpc. The reconstructed power-spectra of the Gaussian variable in the lognormal model turn out to be closer to linear theory as we already expected from previous works. We have restricted ourselves to redshift $z=3$. Future work should be done to extend this study to other redshifts and investigate its use for galaxy redshift surveys.

This work should contribute towards an optimal  cosmological large-scale structure analysis in the quasi-nonlinear regime.}

\section*{Acknowledgments}

FSK thanks Raul E. Angulo for inestimable discussions and for kindly providing his large-scale N-body simulation for this study. Also warm thanks for many encouraging discussions to Jens Jasche, in particular the long days working together on the sampling schemes. FSK thanks also Mike Hobson for the discussion on applying the Hamiltonian sampling scheme to large-scale structure reconstructions back in 2007. We are indebted to the {\it Intra-European Marie Curie fellowship} with project number 221783 and acronym {\it MCMCLYMAN} for supporting this project. Also thanks to the  {\it The Cluster of Excellence for Fundamental Physics on the Origin and Structure of the Universe} for supporting the final stage of this project. Special thanks to Joseph Mohr for support and constructive criticism.  The authors thank Volker Springel for providing his simulations for the matter statistics analysis. We finally thank Antonella Maselli for encouraging discussions on the IGM.

{\small
\bibliographystyle{mn2e}
\bibliography{lit}
}

\appendix
{
\section{Calculation of the mean lognormal field from the linear density field}
\label{sec:mu}

The lognormal field \citep[see][]{1991MNRAS.248....1C} gives an estimate of the linear-Gaussian density field \citep[see e.g. ][]{viel,simoorg,2009ApJ...698L..90N}. 
\be
\mbi  s\equiv\ln(1+\mbi\delta_{\rm M})-\mbi\mu_s\,.
\ee
The mean field $\mbi\mu_s$ is the ensemble average of $\ln(1+\mbi\delta_{\rm M})$ by definition:
\be
\langle \mbi s\rangle=\langle\ln(1+\mbi\delta_{\rm M})\rangle-\mbi\mu_s=0\,.
\ee
We assume here that the fields are {\it fair samples}, i.e. $\langle\mbi\delta_{\rm M}\rangle=0$ and $\langle \mbi s\rangle=0$.
One can find that  the field $\mbi\mu_s$ can be expressed as a function of the linear field $ \mbi s$ only:
\ba
\mbi\delta_{\rm M}&=&\exp(\mbi s+\mbi\mu_s)-1\nonumber\\
\langle\mbi\delta_{\rm M}\rangle&=&\langle\exp(\mbi s+\mbi\mu_s)\rangle-1=0\nonumber\\
\exp(\mbi\mu_s)&=&\frac{1}{\langle\exp(\mbi s)\rangle}\nonumber\\
\mbi\mu_s&=&-\ln(\langle\exp(\mbi s)\rangle)\,.
\ea
In practice the mean field is calculated in the following way
\be
\mu_s=-\ln\left(\frac{\sum_i\exp( s_{{i}})}{N_{\rm c}}\right)\,.
\ee
Note that to ensure that the field $\mbi s$ has zero mean, we have to impose that the zeroth mode of its power-spectrum vanishes.

\section{Relation between the density field in real- and redshift-space in linear Lagrangian perturbation theory}
\label{sec:pecvel}

To correct for the redshift distortions in the linear regime we follow the work of \citet[][]{Kaiser-87} and in particular the expressions derived in linear Lagrangian perturbation theory (LPT) \citep[see][]{1970A&A.....5...84Z} by \citet[][]{1990MNRAS.242..428M}.

We can write in Lagrangian perturbation theory the Eulerian coordinates $\mbi x$  as the sum of the Lagrangian coordinates $\mbi q$ and a displacement field $\mbi \Psi$:
\be
\mbi x=\mbi q+\mbi \Psi\,.
\ee
The analogous expression in redshift-space can be written as
\be
\mbi x^z=\mbi q+\mbi \Psi^z\,,
\ee
with the displacement field in redshift-space given by
\be
\mbi \Psi^z=\mbi\Psi+f(\mbi\Psi\cdot\hat{\mbi r})\hat{\mbi r}\,,
\ee
where  $f = {\rm d}\ln D/{\rm d}\ln a$, $H$ is the Hubble constant, $a$ is the scale factor and $\hat{\mbi r}$ is the line-of-sight direction vector.  For flat models with a non-zero
cosmological constant $f\approx\Omega^{5/9}$ \citep[see][]{bouchet1995}, where $\Omega(z)$ is the matter density at a redshift $z$. 
Note that to linear order the velocity is proportional to the displacement field
\be
\mbi v=fHa\,\mbi \Psi\,.
\ee
Imposing conservation of mass one finds 
\be
\delta_{\rm M}=J^{-1}-1\simeq-\nabla\cdot\mbi\Psi\simeq-(fHa)^{-1}\nabla\cdot\mbi v\,,
\ee
and 
\be
\delta_{\rm M}^z=J_z^{-1}-1\simeq-\nabla\cdot\mbi\Psi^z\simeq\delta_{\rm M}-f\nabla\cdot(\mbi\Psi\cdot\hat{\mbi r})\hat{\mbi r}\,,
\ee
with $J=\det(\partial\mbi x/\partial\mbi q)$ and $J_z=\det(\partial\mbi x^z/\partial\mbi q)$ being the Jacobians of the corresponding  coordinate transformations in real- and redshift-space expanded to linear order.
We can then write the relation between the overdensity field in real- and in redshift-space as
\be
\delta_{\rm M}^z=\delta_{\rm M}+\delta_v\,,
\ee
with
\be
\delta_v=-f\nabla\cdot(\mbi\Psi\cdot\hat{\mbi r})\hat{\mbi r}=-(Ha)^{-1}\nabla\cdot(\mbi v\cdot\hat{\mbi r})\hat{\mbi r}\,.
\ee
If we consider observational effects like a mask $w$ and some noise $\epsilon$ then we have in real-space
\be
\delta_{\rm M}^{{\rm obs}}=w\delta_{\rm M}+\epsilon\,,
\ee
and correspondingly in redshift-space
\be
\delta_{\rm M}^{{\rm obs,}z}=w_z\delta_{\rm M}^z+\epsilon_z\,.
\ee
Assuming that the mask is approximately the same in both spaces
$w=w_z$ \citep[for a discussion on this see][]{kitaura_sdss} we find
\be
\delta_{\rm M}^{{\rm obs,}z}=w\delta_{\rm M}+\epsilon+w\delta_v+\epsilon_v\,,
\ee
with
\be
\epsilon_v=\epsilon_z-\epsilon \,.
\ee
We can then define
\be
\delta^{{\rm obs}}_v=w\delta_v+\epsilon_v\,,
\ee
from which following relation holds to linear order 
\be
\delta_{\rm M}^{{\rm obs,}z}=\delta_{\rm M}^{{\rm obs}}+\delta^{{\rm obs}}_v\,.
\ee
}

\section{Likelihoods}
\label{sec:like}

{Here we present the likelihoods which can be used in our framework slightly modified with respect to the ones derived in \citet[][]{kitaura_log} as we want to estimate the signal $\mbi s=\ln(1+\mbi\delta_{\rm M})-\mbi\mu_s$.}

\subsubsection{Poissonian/Gamma likelihood}

The Poissonian  distribution function \citep[or Gamma function, see][and references there-in]{kitaura} considers the noise to be independent from cell to cell and can include a binomial model for the completeness \citep[see][]{kitaura_sdss}. In this way the noise is proportional to the square root of the completeness in each cell. {Note however that the peculiar velocities will introduce a correlation which we are treating separately within the Gibbs-sampling scheme (see \S \ref{sec:joint}).}
The log-likelihood {with mean $\lambda_i=w_i\overline{\rho}(1+\delta_{{\rm M},i})$} reads \citep[see][]{kitaura_log}:
\ba
\label{eq:likepois}
\lefteqn{-\ln{\cal L}(\mbi d|\mbi s)=}\\
&&\hspace{-0.2cm}\sum_i\overline{\rho}w_i\exp(s_i+\mu_s)-\rho_i\ln(\overline{N}w_i\exp(s_i+\mu_s))+\Gamma(1+\rho_i) \nonumber\,,
\ea
with its gradient being: 
\be
-\frac{\partial\ln{\cal L}(\mbi d|\mbi s)}{\partial s_l}=\overline{\rho}w_l\exp(s_l+\mu_s)-\rho_l \,.
\ee
{Please note, that the density in each cell $\rho_i$ (number counts in the discrete case) is a particular realization of the mean $\lambda_i$ and $\overline{\rho}$ represents the mean density (mean number counts in the discrete case). 
The Gamma function is replaced by a factorial in the discrete case.} Nevertheless, we show below (see Sec.~\ref{sec:MCMC}) that this term has not to be computed. {The relation between the density $\rho$ and the observed overdensity $\delta^{\rm obs}_{\rm M}$ is given by $\delta^{\rm obs}_{{\rm M},i}=\rho_i/\overline{\rho}-w_i$ \citep[see][]{kitaura_sdss}.
}

\subsubsection{Gaussian likelihood}
\label{sec:gauss}

If one wants to include specific errors in each cell {or even a complex correlation in the noise covariance matrix $\mat N$} one can use the Gaussian likelihood.
The log-likelihood for a Gaussian distribution is given by
\be
-\ln{\cal L}(\mbi d|\mbi s)=\frac{1}{2}{\rm ln}\left(\left(2\pi\right)^{N_{\rm c}} {\rm det}(\mat N)\right)+\frac{1}{2}\mbi\epsilon^\dagger\mat N^{-1}\mbi\epsilon\, ,
\ee
with the noise being defined as: $\mbi\epsilon\equiv\mat R\mbi \delta_{\rm M}-\mbi d$. The quantity which needs to be defined is the noise covariance matrix {which in its simplest form is given by a diagonal matrix}: $N_{ij}=\sigma^2_{i}\delta_{ij}^{\rm K}$, i.~e.~the effective variance in each cell: $\sigma^2_{i}$. The variance {in such a model} should be computed summing up the contributions of all the errors in the 1D reconstruction corresponding to a given cell.
We refer to \citet{pichon} to model a more complex noise covariance matrix.

\section{Hamiltonian sampling}
\label{sec:MCMC}

In this appendix we revise the Hamiltonian sampling approach used to sample the conditional matter field and point out the difficulty in its direct use to power-spectrum estimation which justifies the approach presented in \S \ref{sec:joint}. Please also note, that the required expressions slightly change with respect to \citet{kitaura_log,jasche_hamil} as we want to extract here the signal $\mbi s=\ln(1+\mbi\delta_{\rm M})-\mbi\mu_s$ (see appendix \ref{sec:mu}).

The product of the prior and the likelihood is proportional to the posterior distribution function by Bayes theorem:
\begin{equation}
{P}(\mbi s|\mbi d,\mbi c)=\frac{{P}(\mbi s|\mbi c){\mathcal L}(\mbi d|\mbi s)}{\int {\rm d} \mbi s\,{P}(\mbi s|\mbi c){\mathcal L}(\mbi d|\mbi s)}\, {,}
\end{equation}
where the normalization is the so-called evidence.
For many applications one can ignore the evidence. This is the case when one wants to find the maximum a posteriori solution or when one wants to sample the full posterior with a fixed power-spectrum. For all these cases the evidence can be considered just a constant.

Let us define the so-called {\it potential energy} from the posterior distribution ${P}(\mbi s|\mbi d,\mbi c)$:
\begin{equation}
\label{eq:energy}
 E (\mbi s)\equiv-{\rm ln}({P}(\mbi s|\mbi d,\mbi c)) \, .
\end{equation}
For a lognormal  prior and an arbitrary likelihood we have:
\begin{eqnarray}
\label{eq:energy2}
\lefteqn{ E (\mbi s) = \frac{1}{2} {\rm ln}\left(\left(2\pi\right)^{N_{\rm c}} {\rm det}(\mat S)\right)} \\
& &\hspace{.0cm} + \frac{1}{2} \mbi s^\dagger \mat S^{-1} \mbi s
\nonumber \\
&  &\hspace{.0cm}-{\rm ln}\left({\mathcal L}\left(\mbi d|\mbi s\right)\right)+{\rm ln}\left(\int {\rm d}\mbi s\,{P}(\mbi s|\mbi c){\mathcal L}(\mbi d|\mbi s)\right) \,{.} \nonumber 
\end{eqnarray}
Please note that the terms including the determinant of the correlation function and the evidence do not depend on {particular realizations of} the  signal {but on an ensemble of signals. This permits us to consider them as constants when calculating its gradient. Hence, the} gradient of the {\it potential energy} function with respect to the signal is given by:
\begin{equation}
\label{eq:gradpot}
\frac{\partial   E (\mbi s)}{\partial s_l}= \sum_{j} S^{-1}_{lj} s_j - \frac{\partial}{\partial s_l}{\rm ln}\left({\mathcal L}\left(\mbi d|\mbi s\right)\right)\,{.}
\end{equation}
The full posterior  distribution can be obtained by sampling with a Markov Chain Monte Carlo method.
 Given an  analytic expression for the posterior, from which appropriate derivatives can be calculated, one can use Dynamical and Hybrid Monte Carlo methods to sample its whole distribution \citep[see the review by][]{neal1993}.
In particular the Hamiltonian sampling method introduced by \citet[][]{duane} has been proved to efficiently deal with  linear and nonlinear problems in high dimensional spaces\footnote{Each cell $i$ implies a statistical dimension. The problem we face here has therefore $N_\cc$ dimensions.} \citep[see][]{2008MNRAS.389.1284T,jasche_hamil,jasche_sdss}.
The Hamiltonian sampling is a stochastic dynamics method based on a thermodynamical analogy. In this model a {\it force} defined by the gradient of the {\it potential energy} with respect to the system's configuration coordinates changes its configuration through the {\it momenta}. The system visits different configuration states with a frequency given by its {\it canonical} distribution when it is exposed to a {\it heat bath}. This   sampling process   can be modeled through random realizations within the Hamiltonian scheme.

Let us define the Hamiltonian {\it total energy} function:
\begin{equation}
  \label{eq:ham}
  H(\mbi s,\mbi p)=K(\mbi p)+E(\mbi s)\,,
\end{equation}
with a {\it kinetic energy} term constructed on the nuisance parameters given by the {\it momenta} $\mbi p$ and {\it mass} variance $\mat M$: 
\begin{equation}
  \label{eq:kin}
  K(\mbi p)\equiv\frac{1}{2}\sum_{ij}p_iM_{ij}^{-1}p_j\,.
\end{equation}
Note that the {\it potential energy} $E(\mbi s)$ was already defined in Eq.~(\ref{eq:energy}).

Let us use the following compact notation: $P(\mbi p)\equiv P(\mbi p|\mat M)$, $P(\mbi s)\equiv P(\mbi s|\mbi d,\mbi c)$ and $P(\mbi s,\mbi p)\equiv P(\mbi s,\mbi p|\mbi d,\mbi c,\mat M)$. The {\it canonical} distribution function defined by the Hamiltonian (or the joint distribution function of the signal and {\it momenta}) is then given by:
\begin{eqnarray}
  \label{eq:joint}
  P(\mbi s,\mbi p)&=&\frac{1}{Z_H}\exp(-H(\mbi s,\mbi p)) \nonumber\\
  &=&\left[\frac{1}{Z_K}\exp(-K(\mbi p))\right]\left[\frac{1}{Z_E}\exp(-E(\mbi s))\right] \nonumber\\
  &=&P(\mbi p)P(\mbi s) \,,
\end{eqnarray}
with $Z_H$, $Z_K$ and $Z_E$ being the {\it partition} functions so that the probability distribution functions are normalized to one. In particular, the normalization of the Gaussian distribution for the {\it momenta} is represented by the {\it kinetic partition} function $Z_K$.  The Hamiltonian sampling technique does not require the terms which are independent of the configuration coordinates as we will show below.

From Eq.~(\ref{eq:joint}) it can be noticed that in case we have a method to sample
from the joint distribution function  $P(\mbi s,\mbi p)$, marginalizing over the
momenta we can in fact, sample  the posterior $P(\mbi s)$.

The Hamiltonian dynamics provides such a method. We can define a dynamics on {\it phase-space} with the introduction of a {\it time} parameter $t$. The Hamiltonian equations of motion are given by:
\begin{eqnarray}
  \label{eq:EoM1}
  \frac{ds_i}{dt}& =& \frac{\partial H}{\partial p_i}=\sum_j M^{-1}_{ij} p_j\, ,\\
  \label{eq:EoM2}
  \frac{dp_i}{dt}& =& - \frac{\partial H}{\partial s_i} = - \frac{\partial E(s)}{\partial s_i}\, .
\end{eqnarray}
To sample the posterior one has to solve these equations for randomly drawn {\it momenta} according to the kinetic term defined by Eq.~(\ref{eq:kin}). This is done by drawing  Gaussian samples with a variance given by the {\it mass} $\mat M$ which can tune the efficiency of the sampler \citep[see][]{jasche_hamil}. The marginalization over the {\it momenta} occurs by drawing new {\it momenta} for each Hamiltonian step disregarding the ones of the previous step.

It is not possible to follow the dynamics exactly, as one has to use a discretized version of the equations of motion. It is convenient to use the {\it leapfrog} scheme which has the properties of being {\it time}-reversible and conserve {\it phase-space} volume being necessary conditions to ensure {\it ergodicity}:
\begin{eqnarray}
  \label{eq:leap1}
  p_i\left(t+\frac{\epsilon}{2}\right) &=& p_i(t) -\frac{\epsilon}{2} \left .\frac{\partial  E(\mbi s)}{\partial s_l} \right |_{s_i(t)} \, , \\
\label{eq:leap2}
  s_i\left(t+\epsilon \right) &=& s_i(t) +\epsilon\sum_jM_{ij}^{-1}\,p_j\left(t+\frac{\epsilon}{2}\right)  \, , \\
  p_i\left(t+\epsilon\right) &=& p_i\left(t+\frac{\epsilon}{2}\right) -\frac{\epsilon}{2} \left .\frac{\partial  E(\mbi s)}{\partial s_l} \right |_{s_i\left(t+\epsilon \right)} \, . 
\end{eqnarray}
The dynamics of this system are followed for a period of {\it time} $\Delta \tau$, with a value of $\epsilon$ small enough to give acceptable errors and for $N_\tau=\Delta\tau/\epsilon$ iterations. In practice $\epsilon$ and $N_\tau$ are randomly drawn from a uniform distribution to avoid resonant trajectories \citep[see][]{neal1993}.

The solution of the equations of motion will move the system from an initial state  \((\mbi s,\mbi p)\)  to a final state \((\mbi s',\mbi p')\) after each sampling step.
Although the Hamiltonian equations of motion are energy conserving, our approximate solution is not. Moreover, the starting guess will not be drawn from the correct distribution and a {\it burn-in} phase will be needed. For these reasons a Metropolis-Hastings acceptance step has to be introduced in which the new {\it phase-space} state \((\mbi s',\mbi p')\) is accepted with probability:
\begin{equation}
\label{eq:acc}
{P}_A = {\rm min}\left[1,{\rm exp}(-\left(\delta H)\right)\right]\, ,
\end{equation}
with $\delta H\equiv H(\mbi s',\mbi p')-H(\mbi s,\mbi p)$.

The Hamiltonian sampling technique cannot be easily extended to jointly sample the density and the power-spectrum. 
As we can see from Eq.~(\ref{eq:acc}) the Hamiltonian sampler does not require the evaluation of the full joint posterior $H(\mbi s,\mbi p)$, but just the difference between two subsequent states $\delta H$. 
Such a calculation is simple when the cosmological parameters are fixed, but becomes extremely complicated when one samples them within the same Markov Chain. 
The reason being that calculation of the determinant of the covariance and the evidence {would become} necessary in each Hamiltonian step (see Eq.~\ref{eq:energy2}).
 {This problem can be overcome with the Gibbs-sampling scheme given that the conditional probability distribution functions are sampled in a consistent way as we show in \S \ref{sec:joint}. }

\end{document}